\documentclass[12pt,preprint]{aastex}

\usepackage{graphicx}
\usepackage{epstopdf}
\usepackage{xcolor}

\newcommand{\et}{et al.}
\newcommand{\solar}{\ifmmode_{\sun}\;\else$_{\sun}\;$\fi}
\newcommand{\kms}{km s$^{-1}$}
\newcommand{\HI}{H$\,${\sc i}}
\newcommand{\HII}{H$\,${\sc ii}}
\newcommand{\ha}{H$\alpha$}

\newcommand{\sfr}{$\log {\rm SFR_D^{FUV}}$}
\newcommand{\oh}{$12+\log {\rm (O/H)}$}
\newcommand{\paramhi}{$\log \Sigma_{\rm HI}^0$}
\newcommand{\paramr}{$R_{0,HI}$}
\newcommand{\paramn}{$n_{HI}$}
\newcommand{\paramvc}{$V_c$}
\newcommand{\paramrt}{$R_t$}
\newcommand{\paramgamma}{$\gamma$}

\newcommand{\unithi}{M\solar\ pc$^{-2}$}
\newcommand{\dirr}{dIrr}
\newcommand{\rd}{$R_{\rm D}$}
\newcommand{\rbr}{$R_{\rm Br}$}
\newcommand{\rha}{$R_{\rm H\alpha}$}
\newcommand{\rfuvknot}{$R_{\rm FUV knot}$}
\newcommand{\logfuvrat}{$\rm \log FUV_{1R_D}/FUV_{1-3R_D}$}
\newcommand{\rbar}{$R_{\rm Bar\ end}$}
\newcommand{\paramfuv}{$\log \mu_{\rm FUV}^0$}
\newcommand{\paramrfuv}{$R_{0,FUV}$}
\newcommand{\paramnfuv}{$n_{FUV}$}
\newcommand{\rfifty}{$R_{50,HI}$}
\newcommand{\rninety}{$R_{90,HI}$}
\newcommand{\rrat}{$(R_{50}/R_{90})_{HI}$}

\begin{document}

\title{Relationships between the Stellar, Gaseous,
and Star Formation Disks in LITTLE THINGS Dwarf Irregular Galaxies: Indirect
Evidence for Substantial Fractions of Dark Molecular Gas}

\author{
Deidre A.\ Hunter\altaffilmark{1},
Bruce G.\ Elmegreen\altaffilmark{2},
Esther Goldberger\altaffilmark{1,3},
Hannah Taylor\altaffilmark{1,4},
Anton I.\ Ermakov\altaffilmark{1,4,5},
Kimberly A.\ Herrmann\altaffilmark{6},
Se-Heon Oh\altaffilmark{7},
Bradley Malko\altaffilmark{1,8},
Brian Barandi\altaffilmark{1,8},
Ryan Jundt\altaffilmark{1,8}
}

\altaffiltext{1}{Lowell Observatory, 1400 West Mars Hill Road, Flagstaff, AZ 86001, USA}
\altaffiltext{2}{IBM T.\ J.\ Watson Research Center, PO Box 218, Yorktown Heights, New York}
\altaffiltext{3}{Department of Physics, Massachusetts Institute of Technology, 77 Massachusetts Ave., 54-918, Cambridge, MA 02142, USA}
\altaffiltext{4}{Department of Earth, Atmospheric, and Planetary Sciences, Massachusetts Institute of Technology, 77 Massachusetts Avenue, Cambridge, MA 02139-4307, USA}
\altaffiltext{5}{Current affiliation Earth and Planetary Science Department, University of California-Berkeley, Berkeley, CA 94709, USA}
\altaffiltext{6}{Department of Physics, Pennsylvania State University Mont Alto, 1 Campus Drive, Mont Alto, PA 17237, USA}
\altaffiltext{7}{Department of Physics and Astronomy, Sejong University, 209 Neungdong-ro, Gwangjin-gu, Seoul, Republic of Korea}
\altaffiltext{8}{Department of Astronomy and Planetary Sciences, Northern Arizona University, Flagstaff, AZ 86001, USA}

\begin{abstract}
The stellar, gaseous and young stellar disks in the LITTLE THINGS sample of
nearby dIrrs are fitted with functions to search for correlations between
the parameters. We find that the \HI\ radial profiles are generally flatter in the
center and fall faster in the outer regions than the $V$-band profiles, while young
stars are more centrally concentrated, especially if the \HI\ is more centrally
flat. This pattern suggests that the \HI\ is turning into molecules in the center
and the molecular clouds are forming stars and FUV. 
A model that assumes the molecular surface density is proportional to the total gas
surface density to a power of 1.5 or 2, in analogy with the Kennicutt-Schmidt relation,
reproduces the relationship between the ratio of the visible to the \HI\ scale length and
the \HI\ Sersic index. The molecular fraction is estimated as a function of radius
for each galaxy by converting the FUV to a molecular surface density using conventional
calibrations. The average molecular fraction inside $3R_{\rm D}$ is $23\pm17$\%.
However, the break in the stellar surface brightness profile 
has no unified tracer related to star formation.
\end{abstract}


\section{Introduction} \label{sec-intro}

Dwarf irregular (\dirr) galaxies contain stellar populations and most also have on-going star formation.
Yet, their atomic gas surface densities, even in their centers, are low compared to those in the central regions of spirals.
In fact the gas surface densities in \dirr\ galaxies and the outer parts of spirals
are lower than the threshold necessary for gravitational instabilities to make clouds
that can then go on to make stars as is believed to happen in the central regions of spirals 
\citep{toomre64,kennicutt89,hunter98,bigiel10,barnes12,kennicutt89,eh15}.
Furthermore, \dirr\ galaxies have extended stellar disks that have
been traced up to 12 disk scale lengths, \rd,
\citep[e.g.][]{saha10,sanna10,outer11,bellazzini14}, and
young stars are found in the far outer parts of \dirr\ and spiral galaxies \citep{thilker05,outerfuv}.
Thus, star formation appears to be taking place in extreme environments of sub-threshold gas densities.
In addition,
star formation appears to have proceeded ``outside-in'' in dIrrs \citep{zhang12,gallart08,meschin14,pan15}.
By contrast, spiral stellar disks are observed to grow from ``inside-out'' \citep[e.g.,][]{williams09}.
The cause of this difference in disk formation is also not understood.

\HI\ gas extends well beyond the bright stellar part of the galaxy in both
spirals \citep[e.g.][]{warmels86,broeils92,rao93,hulst93} and
dwarfs \citep[e.g.][]{hunter85,meurer96} with an unusual concentration of
\HI\ towards the centers in Blue Compact Dwarfs \citep[e.g.][]{zee98,simpson00}.
The ratio of the
mass in stars to mass in atomic gas drops steadily with radius in \dirr\ galaxies \citep[see, for
example, Figure 4 in][]{hunter08}. Dwarfs are usually gas dominated even in the
centers, but become more so with radius. This implies a steady decrease in the
efficiency of conversion of atomic gas into stars with distance from the center of
the galaxy \citep[see, for example,][]{leroy08}.

Since stars form from dense gas clouds that presumably form from the
general atomic interstellar medium (ISM) in a galaxy, we expect there to be a
relationship between gas density and cloud-forming instabilities \citep[e.g.][]{toomre64} and, 
hence, between gas density and the formation of stars \citep{goldreich65}.
Yet, the apparent wide variety of ways the gas surface density falls off with
radius in \dirr\ galaxies is striking. Some profiles are relatively flat, some drop precipitously, and
others decrease steadily. This variety seems to be far greater than the uniformly
exponential surface brightness profiles seen in the stellar disks of most \dirr\
systems. 
Empirical relationships show a general correspondence between gas and star formation
\citep[see, for example,][]{bigiel08,bigiel10}. Yet, the physical connection
between large-scale gas distributions and the formation of new stars is still
illusive and empirically a large range in star formation rates (SFRs) can be found
at a given surface density of the atomic gas. What then is the role of the gas in
determining the nature of the stellar disk? This question is particularly compelling
in outer stellar disks where we see young stars but the gas densities are especially
low.

However, there is a further complication in understanding how the gas and stars are related: 
most stellar disk surface brightnesses do not drop off with radius at a
single rate. Most, both spiral and \dirr, show a break in their stellar exponential disks. 
The stellar surface brightness profile drops off exponentially, and then at
the break radius \rbr\ it either drops more steeply \citep[Type II break,][]{freeman70} or drops less
steeply \citep[the less common Type III break, which could be the signature of a lopsided
disk,][]{erwin05,watkins19}. 
A Type I disk has no break \citep{freeman70}. 
In spiral galaxies, the break is not as apparent in the mass surface density profiles as in the
stellar surface brightness profiles \citep{bakos08} and this could be related
to the potential for spiral arms to scatter inner disk stars to the outer regions \citep{bournaud07,roskar08}.
On the other hand, scattering
is less effective in \dirr\ galaxies \citep{struck17}, which do not have spiral arms,
and the break is also seen in the stellar mass surface density profiles of these galaxies \citep{herrmann16}.
Star formation processes could also 
vary with radius, including the ability to form molecules, which should be more
prevalent in the inner regions of dwarfs than the outer regions
\citep[e.g.,][]{hunter19a}, which could potentially lead to a break in the stellar profile.
Alternatively, \citet{andersen00} suggest that
self-regulated evolution within a confining dark halo leads to exponential density
profiles that are somewhat flatter in the central regions.

To explore the factors at play in determining the structure of the stellar disk in
dwarf galaxies, we have parameterized the radial profiles of \HI\ mass surface densities,
\HI\ rotation, stellar surface brightness, and star formation activities of a sample of nearby
dwarf galaxies that are part of LITTLE THINGS \citep[Local Irregulars That Trace
Luminosity Extremes, The \HI\ Nearby Galaxy Survey;][]{hunter12}. We compare the
characteristics of the gas and star formation activity with those of the stellar
disk, looking for correlations that could be clues to processes that shape the
stellar disk. We also look for hidden H$_2$ in the form of missing gas
connected with star formation.

\section{Data}
\label{sec-data}

LITTLE THINGS\footnote[9]{The original VLA survey was funded in part by the National Science Foundation through grants AST-0707563, AST-0707426, AST-0707468, and AST-0707835 to US-based LITTLE THINGS team members and supported with generous technical and logistical support from the National Radio Astronomy Observatory.}
is a multi-wavelength survey aimed at
determining what drives star formation in dwarf galaxies \citep{hunter12}.
The LITTLE THINGS sample includes 37 \dirr\ galaxies and
4 Blue Compact Dwarf (BCD) galaxies, and is centered around \HI-emission
data obtained with the
National Science Foundation's Karl G.\ Jansky Very Large Array
(VLA\footnote[10]{The VLA is a facility of the National Radio Astronomy Observatory.
The National Radio Astronomy Observatory is a facility of the National Science Foundation operated under cooperative agreement
by Associated Universities, Inc.}).
The \HI-line data are characterized by high sensitivity ($\leq1.1$
mJy beam$^{-1}$ per channel), high spectral resolution (1.3 or 2.6 \kms),
and high angular resolution (typically 6\arcsec). The LITTLE THINGS
sample contains dwarf galaxies that are relatively nearby ($\leq$10.3 Mpc;
6\arcsec\ is $\leq$300 pc), contain gas so they have the potential for star formation,
and are not companions to larger galaxies. The sample was also chosen
to cover a large range in dwarf galactic properties such as
SFR and absolute magnitude.

The LITTLE THINGS ancillary data include far-ultraviolet (FUV) images
obtained with the NASA {\it Galaxy Evolution Explorer} satellite
\citep[{\it GALEX\footnote[11]{{\it GALEX} was operated
for NASA by the California Institute of Technology under NASA contract NAS5-98034.}};][]{galex}
to trace star formation over the past 200 Myr.
These data give us integrated SFRs \citep{hunter10} and the radius at which we found the furthest out FUV knot \rfuvknot\ in each galaxy
\citep{outerfuv}.
So that galaxies can be compared, the SFRs are normalized to the area within one disk scale length and are, technically,
SFR surface densities, although star formation is usually found beyond 1\rd.
LITTLE THINGS \ha\ images give us the SFR over the past 10 Myr and the radius of the furthest out \HII\ region \rha\ \citep{he04}.
Surface photometry of $UBVJHK$ images was used by \citet{herrmann13} and \citet{herrmann16} to investigate the breaks
in stellar surface brightness profiles, the radius at which there is a sudden change in the slope of the exponential decline.
Here we use the break radius \rbr\ and disk scale length \rd\ determined from the $V$-band image.
We also use the integrated galactic luminosities from \citet{he06}.

The galaxy sample and characteristics that we use here are given in Table \ref{tab-gal}.
In some plots, we distinguish between those dIrrs that are classified as Magellanic irregulars (dIm)
and those that are classified as BCDs (Haro 29, Haro 36, Mrk 178, VIIZw403).

\begin{deluxetable}{lcccccccccc}
\tabletypesize{\tiny}
\tablecaption{The Galaxy Sample \label{tab-gal}}
\tablewidth{0pt}
\tablehead{
\colhead{} & \colhead{D\tablenotemark{a}} & \colhead{M$_V$} & \colhead{$R_{\rm H\alpha}$\tablenotemark{b}}
& \colhead{\rfuvknot\tablenotemark{c}}
& \colhead{$R_{\rm D}$\tablenotemark{d}}  & \colhead{$R_{\rm Br}$\tablenotemark{e}}
& \colhead{\sfr\tablenotemark{f}}
&\colhead{$C_{31}$\tablenotemark{g}}
& \colhead{$\rm \log FUV_{1R_D}/FUV_{1-3R_D}$\tablenotemark{h}} \\
\colhead{Galaxy} &  \colhead{(Mpc)} & \colhead{} & \colhead{(kpc)}  & \colhead{(kpc)} & \colhead{(kpc)} & \colhead{(kpc)}
& \colhead{(M\solar\ yr$^{-1}$ kpc$^{-2}$)} & \colhead{} & \colhead{} & \colhead{} \\
}
\startdata
CVnIdwA  &  $3.6\pm0.08$  & $-12.37\pm0.09$ &      0.69 & 0.49$\pm$0.03 & 0.25$\pm$0.12 & 0.56$\pm$0.49 &   $-1.77\pm0.04$ &      2.53 & $-0.14\pm0.06$ \\
DDO 43    &  $7.8\pm0.8$  & $-15.06\pm0.22$ &      2.36 & 1.93$\pm$0.08 & 0.87$\pm$0.10 & 1.46$\pm$0.53 &     $-2.20\pm0.04$ & \nodata & $0.60\pm0.06$ \\
DDO 46    &  $6.1\pm0.4$  & $-14.67\pm0.16$ &      1.51 & 3.02$\pm$0.06 & 1.13$\pm$0.05 & 1.27$\pm$0.18 &     $-2.45\pm0.04$ & \nodata & $0.92\pm0.06$ \\
DDO 47    &  $5.2\pm0.6$  & $-15.46\pm0.24$ &      5.58 & 5.58$\pm$0.05 & 1.34$\pm$0.05 & \nodata             &     $-2.38\pm0.04$ & \nodata & $0.54\pm0.06$ \\
DDO 50    &  $3.4\pm0.05$  & $-16.61\pm0.03$ & \nodata & 4.86$\pm$0.03 & 1.48$\pm$0.06 & 2.65$\pm$0.27 &   $-1.81\pm0.04$ &      2.45 & $0.51\pm0.06$ \\
DDO 52    & $10.3\pm0.8$ & $-15.45\pm0.17$ &      3.69 & 3.39$\pm$0.10 & 1.26$\pm$0.04 & 2.80$\pm$1.35 &     $-2.53\pm0.04$ &      2.68 & $0.42\pm0.06$ \\
DDO 53    &   $3.6\pm0.05$ & $-13.84\pm0.03$ &      1.25 & 1.19$\pm$0.03 & 0.47$\pm$0.01 & 0.62$\pm$0.09 &   $-1.96\pm0.04$ &      2.10 & $0.68\pm0.06$ \\
DDO 63    &   $3.9\pm0.05$ & $-14.78\pm0.03$ &      2.26 & 2.89$\pm$0.04 & 0.68$\pm$0.01 & 1.31$\pm$0.10 &   $-2.05\pm0.04$ &      2.29 & $0.12\pm0.06$ \\
DDO 69    &   $0.8\pm0.04$ & $-11.67\pm0.11$ &      0.76 & 0.76$\pm$0.01 & 0.19$\pm$0.01 & 0.27$\pm$0.05 &   $-2.22\pm0.04$ &      2.36 & $0.28\pm0.06$ \\
DDO 70    &   $1.3\pm0.07$ & $-14.10\pm0.12$ &      1.23 & 1.34$\pm$0.01 & 0.44$\pm$0.01 & 0.13$\pm$0.07 &  $-2.17\pm0.04$ &      2.77 & $1.03\pm0.06$ \\
DDO 75    &   $1.3\pm0.05$ & $-13.91\pm0.08$ &      1.17 & 1.38$\pm$0.01 & 0.18$\pm$0.01 & 0.71$\pm$0.08 &  $-0.99\pm0.04$ &      2.03 & $-0.12\pm0.06$ \\
DDO 87    &   $7.7\pm0.5$ & $-14.98\pm0.15$ &      3.18 & 4.23$\pm$0.07 & 1.21$\pm$0.02 & 0.99$\pm$0.11 &  $-2.61\pm0.04$ &      2.69 & $0.22\pm0.06$ \\
DDO 101  &   $6.4\pm0.5$ & $-15.01\pm0.16$ &      1.23 & 1.23$\pm$0.06 & 0.97$\pm$0.06 & 1.16$\pm$0.11 &  $-2.84\pm0.04$ &      2.52 & $0.75\pm0.06$ \\
DDO 126  &   $4.9\pm0.5$ & $-14.85\pm0.24$ &      2.84 & 3.37$\pm$0.05 & 0.84$\pm$0.13 & 0.60$\pm$0.05 &   $-2.18\pm0.04$ &      2.58 & $0.57\pm0.06$ \\
DDO 133  &  $3.5\pm0.2$ & $-14.75\pm0.16$ &      2.60 & 2.20$\pm$0.03 & 1.22$\pm$0.04 & 2.25$\pm$0.24 &     $-2.60\pm0.04$ &      2.54 & $0.61\pm0.06$ \\
DDO 154  &   $3.7\pm0.3$ & $-14.19\pm0.16$ &      1.73 & 2.65$\pm$0.04 & 0.48$\pm$0.02 & 0.62$\pm$0.09 &   $-1.77\pm0.04$ &      2.47 & $0.32\pm0.06$ \\
DDO 155  &   $2.2\pm0.4$ & $-12.53\pm0.36$ &      0.67 & \nodata            & 0.15$\pm$0.01 & 0.20$\pm$0.04 & \nodata &      3.06 & \nodata \\
DDO 165   &  $4.6\pm0.4$ & $-15.60\pm0.19$ &      3.16 & \nodata            & 2.24$\pm$0.08 & 1.46$\pm$0.08 & \nodata &      2.30 & \nodata \\
DDO 167   &  $4.2\pm0.5$ & $-12.98\pm0.25$ &      0.81 & 0.70$\pm$0.04 & 0.22$\pm$0.01 & 0.56$\pm$0.11 &   $-1.59\pm0.04$ & \nodata & $0.03\pm0.06$ \\
DDO 168   &  $4.3\pm0.5$ & $-15.72\pm0.25$ &      2.24 & 2.25$\pm$0.04 & 0.83$\pm$0.01 & 0.72$\pm$0.07 &   $-2.06\pm0.04$ &      2.64 & $0.55\pm0.06$ \\
DDO 187   &  $2.2\pm0.07$ & $-12.68\pm0.07$ &      0.30 & 0.42$\pm$0.02 & 0.37$\pm$0.06 & 0.28$\pm$0.05 & $-2.60\pm0.04$ &      2.51 & $1.39\pm0.06$ \\
DDO 210   &  $0.9\pm0.04$ & $-10.88\pm0.10$ & \nodata & 0.29$\pm$0.01 & 0.16$\pm$0.01 & \nodata             & $-2.66\pm0.04$ &      2.63 & $0.87\pm0.06$ \\
DDO 216   &  $1.1\pm0.05$ & $-13.72\pm0.10$ &      0.42 & 0.59$\pm$0.01 & 0.52$\pm$0.01 & 1.77$\pm$0.45 & $-3.17\pm0.04$ &      2.25 & $1.35\pm0.06$ \\
F564-V3    &  $8.7\pm0.7$ & $-13.97\pm0.18$ & \nodata & 1.24$\pm$0.08 & 0.63$\pm$0.09 & 0.73$\pm$0.40 &  $-2.94\pm0.04$ & \nodata & $0.60\pm0.06$ \\
IC 10         &  $0.7\pm0.05$ & $-16.34\pm0.16$ & \nodata & \nodata          & 0.39$\pm$0.01 & 0.30$\pm$0.04 & \nodata & \nodata & \nodata \\
IC 1613     &  $0.7\pm0.05$ & $-14.60\pm0.16$ & \nodata & 1.77$\pm$0.01 & 0.53$\pm$0.02 & 0.71$\pm$0.12 &  $-1.97\pm0.04$ &      2.64 & $0.26\pm0.06$ \\
LGS 3        &  $0.7\pm0.08$ &  $-9.74\pm0.25$ & \nodata & 0.32$\pm$0.01 & 0.16$\pm$0.01 & 0.27$\pm$0.08 &  $-3.75\pm0.04$ &      2.04 & $0.69\pm0.06$ \\
M81dwA     & $3.6\pm0.2$ & $-11.73\pm0.13$ & \nodata & 0.71$\pm$0.03 & 0.27$\pm$0.00 & 0.38$\pm$0.03 &  $-2.30\pm0.04$ &      2.02 & $0.10\pm0.06$ \\
NGC 1569  & $3.4\pm0.2$ & $-18.24\pm0.13$ & \nodata & 1.14$\pm$0.03 & 0.46$\pm$0.02 & 0.85$\pm$0.24 &  $-0.32\pm0.04$ &      3.13 & $1.41\pm0.06$ \\
NGC 2366  & $3.4\pm0.3$ & $-16.79\pm0.20$ &      5.58 & 6.79$\pm$0.03 & 1.91$\pm$0.25 & 2.57$\pm$0.80 &  $-2.04\pm0.04$ &      2.70 & $1.01\pm0.06$ \\
NGC 3738  & $4.9\pm0.5$ & $-17.12\pm0.24$ &      1.48 & 1.21$\pm$0.05 & 0.77$\pm$0.01 & 1.16$\pm$0.20 &   $-1.52\pm0.04$ &      2.95 & $1.76\pm0.06$ \\
NGC 4163  & $2.9\pm0.04$ & $-14.45\pm0.03$ &      0.88 & 0.47$\pm$0.03 & 0.32$\pm$0.00 & 0.71$\pm$0.48 &  $-1.89\pm0.04$ &      2.62 & $1.36\pm0.06$ \\
NGC 4214  & $3.0\pm0.05$ & $-17.63\pm0.04$ & \nodata & 5.46$\pm$0.03 & 0.75$\pm$0.01 & 0.83$\pm$0.14 &  $-1.11\pm0.04$ &      3.09 & $0.96\pm0.06$ \\
Sag DIG     & $1.1\pm0.07$ & $-12.46\pm0.14$ &      0.51 & 0.65$\pm$0.01 & 0.32$\pm$0.05 & 0.57$\pm$0.14 &  $-2.40\pm0.04$ & \nodata & $1.04\pm0.06$ \\
UGC 8508  & $2.6\pm0.1$ & $-13.59\pm0.13$ &      0.79 & \nodata            & 0.23$\pm$0.01 & 0.41$\pm$0.06 & \nodata                  &      2.49 & \nodata \\
WLM           & $1.0\pm0.07$ & $-14.39\pm0.15$ &      1.24 & 2.06$\pm$0.01 & 1.18$\pm$0.24 & 0.83$\pm$0.16 & $-2.78\pm0.04$ &      2.31 & $1.69\pm0.06$ \\
Haro 29      &  $5.8\pm0.3$ & $-14.62\pm0.11$ &      0.96 & 0.86$\pm$0.06 & 0.33$\pm$0.00 & 1.15$\pm$0.26 &  $-1.21\pm0.04$ &      5.29 & $0.86\pm0.06$ \\
Haro 36      &  $9.3\pm0.6$ & $-15.91\pm0.15$ &      1.06 & 1.79$\pm$0.09 & 1.01$\pm$0.00 & 1.16$\pm$0.13 &   $-1.88\pm0.04$ & \nodata & $1.56\pm0.06$ \\
Mrk 178      &  $3.9\pm0.5$ & $-14.12\pm0.26$ &      1.17 & 1.45$\pm$0.04 & 0.19$\pm$0.00 & 0.38$\pm$0.00 &   $-1.17\pm0.04$ &      2.78 & $0.18\pm0.06$ \\
VIIZw403    &  $4.4\pm0.07$ & $-14.27\pm0.04$ &      1.27 & 0.33$\pm$0.04 & 0.53$\pm$0.02 & 1.02$\pm$0.29 &  $-1.80\pm0.04$ &      2.45 & $1.23\pm0.06$ \\
\enddata
\tablenotetext{a}{Distance to the galaxy. References are given by \citet{hunter12}. Uncertainty in the distance
is folded into the uncertainty of M$_V$.}
\tablenotetext{b}{Radius of furthest out detected \HII\ region \rha\ in each galaxy from \citet{he04}.
Galaxies without \HII\ regions or with \HII\ regions
extending beyond the area imaged do not have $R_{H\alpha}$.}
\tablenotetext{c}{Radius of furthest out detected FUV knot \rfuvknot\ in each galaxy from \citet{outerfuv}.
Galaxies without {\it GALEX} images have no value for this radius.}
\tablenotetext{d}{Disk scale length \rd\ determined from the $V$-band image surface photometry from \citet{herrmann13}.
In the case of galaxies with breaks in their surface brightness profiles, we have chosen the scale length that describes the
primary underlying stellar disk.}
\tablenotetext{e}{Break radius \rbr\ where the $V$-band surface brightness profile changes slope given by \citet{herrmann13}.
Galaxies without \rbr\ do not have breaks in their profiles.}
\tablenotetext{f}{SFR measured from the integrated FUV luminosity and normalized to the area within one \rd\ from \citet{hunter10}.
The normalization is independent of the radial extent of the FUV emission in a galaxy.
We assume an uncertainty of 10\%, which is greater than the photometric uncertainty.}
\tablenotetext{g}{A measure of the central concentration of stars - ratio of the radii that
encompass 75\% and 25\% of the total stellar mass, from \citet{zhang12}.
A larger ratio means the stars are more centrally concentrated.}
\tablenotetext{h}{Ratio of
the FUV emission within the inner disk scale length \rd\ and the FUV emission in the annulus from radius 1\rd\ to 3\rd.
A larger ratio means the FUV emission, and hence star formation, is centrally concentrated.
We assume an uncertainty of 10\% in both the numerator and the denominator.}
\end{deluxetable}

\section{Radial Profiles} \label{sec-profiles}

\subsection{\HI\ Surface Density}

\begin{figure}[t!]
\epsscale{.9}
\plottwo{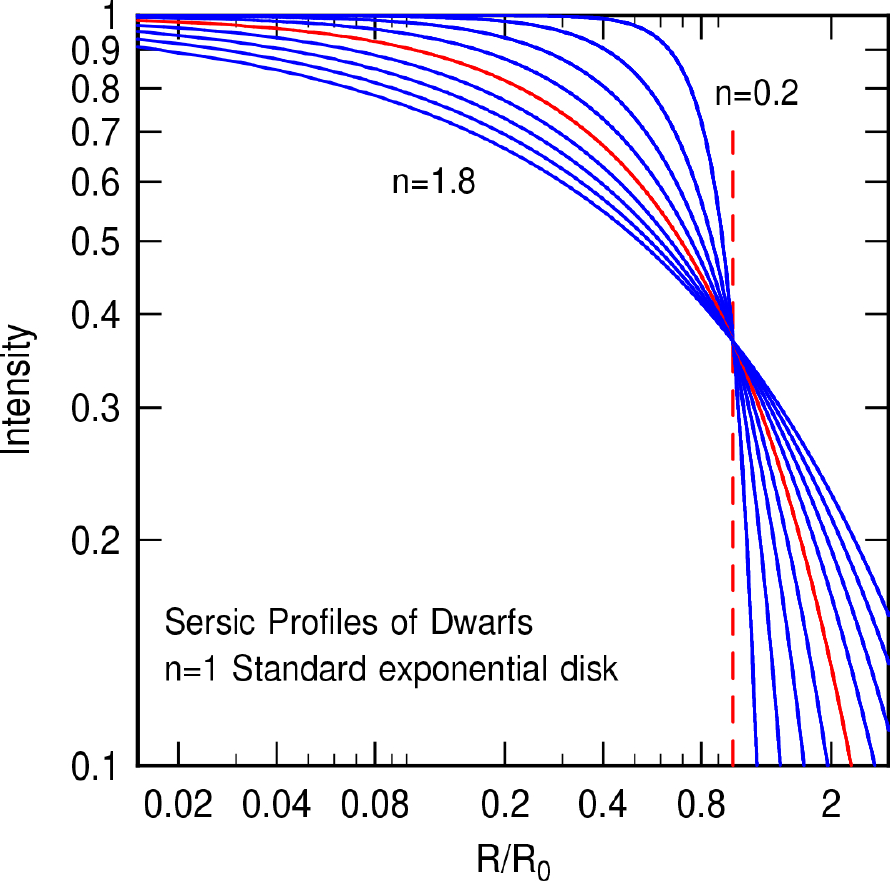}{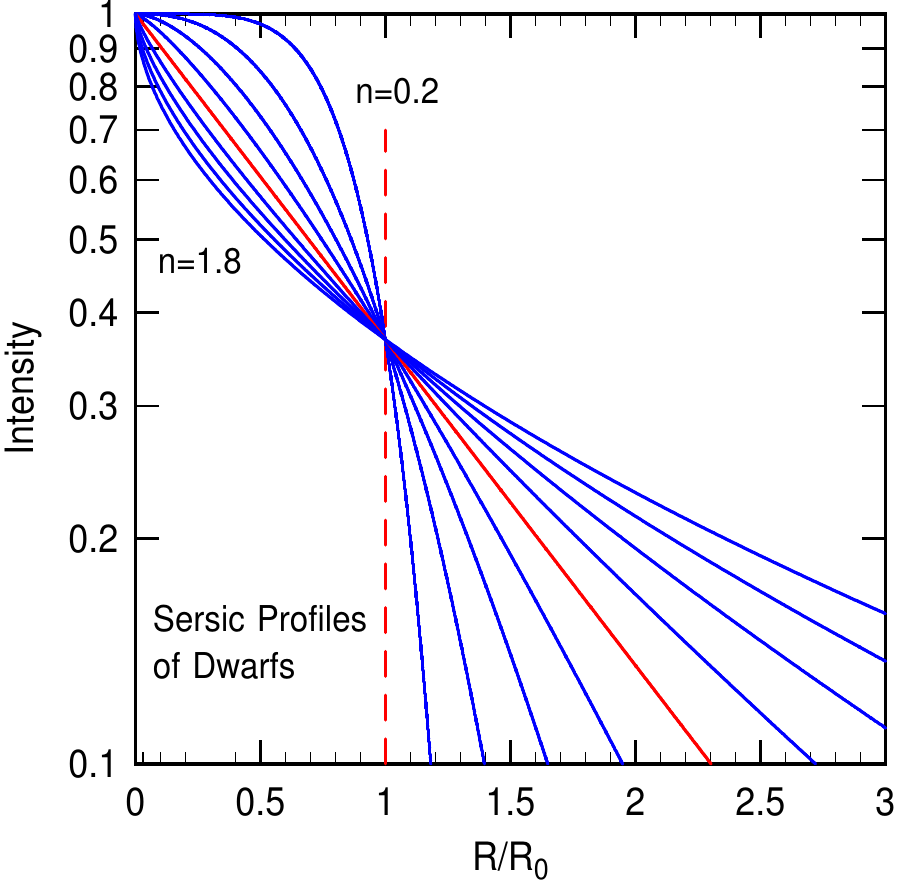}
\caption{Family of Sersic profiles covering the \HI\ fall off in the LITTLE THINGS dwarf galaxy sample
plotted against $\log R$ ({\it left}) and linear $R$ ({\it right}).
An $n=1$ curve (red solid line)
is a standard exponential disk, and $n=4$ would be an
ellipsoidal system \citep[][]{devauc79}. The vertical dashed line marks $R/R_0=1$.
\label{fig-sersic}}
\end{figure}

\begin{figure}
\epsscale{1.}
\plotone{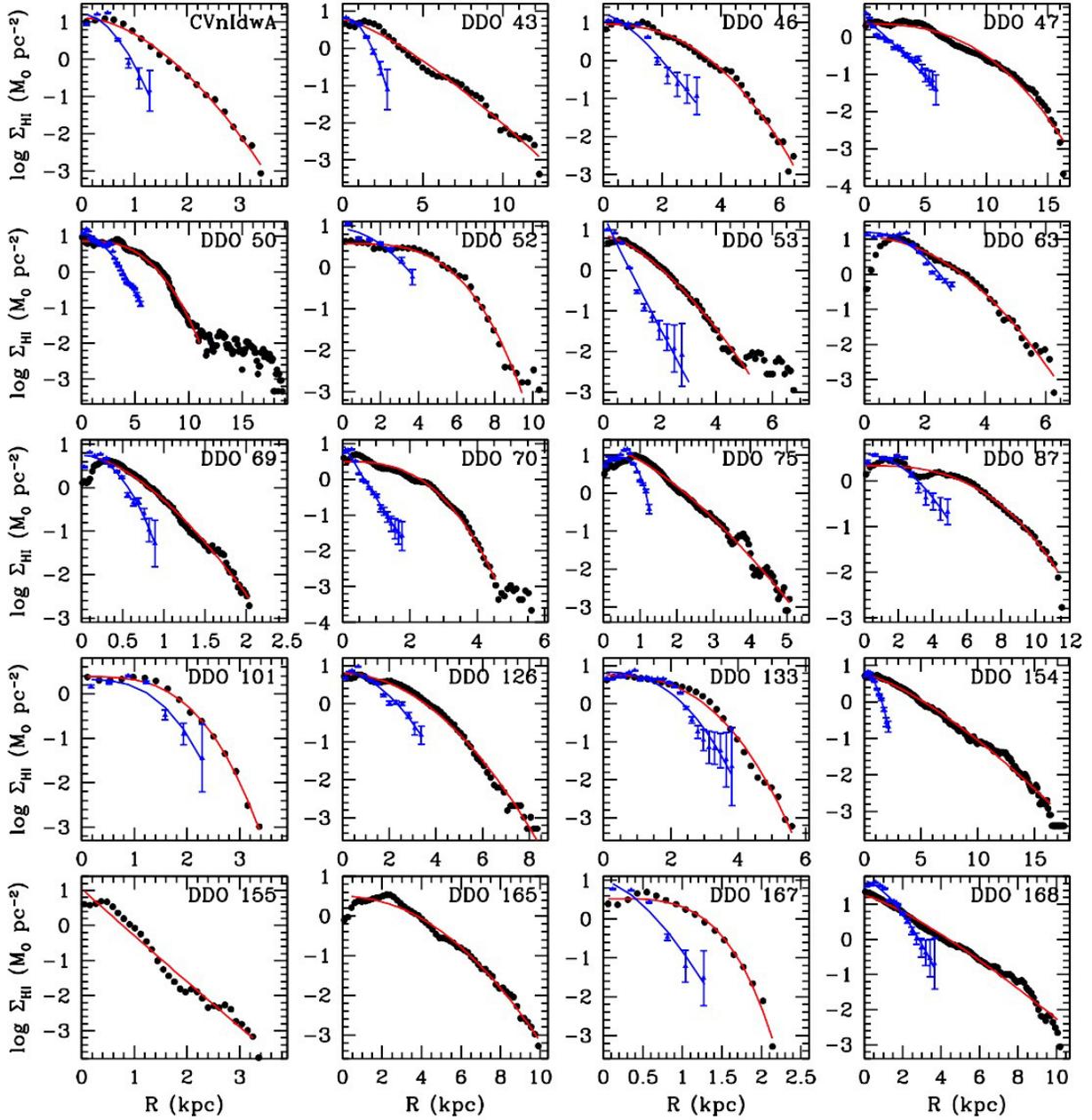}
\caption{\HI\ surface density profiles from \citet{hunter12}.
The red lines are the Sersic fits to the \HI\ profiles.
The blue triangles are FUV surface brightness profiles plotted with the same
logarithmic interval as for the \HI\ \citep{hunter10,zhang12}. Blue curves are
Sersic fits to the FUV. DDO 155, DDO 165, IC 10, and UGC 8508 do not have FUV data.
\label{fig-sersicprof}}
\end{figure}

The \HI\ surface density profiles of the LITTLE THINGS dwarfs are described and shown by \citet{hunter12}.
We performed a multi-variable least squares fit of a
\citet{sersic82} profile to the gas distributions measured from
velocity-integrated {\sc robust}-weighted maps. The Sersic profile, as used here, is
\begin{equation}
I(R) = I_0 e^{-(R/R_0)^{1/n}}.
\end{equation}
For our situation, this can be re-written as
\begin{equation}
\log \Sigma_{\rm HI}(R) = \log \Sigma_{\rm HI}^0 -0.434(R/R_{0,HI})^{1/n_{HI}},
\end{equation}
so that the \HI\ surface density profiles are defined by three parameters:
$\log \Sigma_{\rm HI}^0$, the logarithm of the extrapolated central surface gas density in units of $M_\odot$ pc$^{-2}$;
\paramr, a characteristic radius;
and \paramn, an index that controls the curvature of the profile.
$n$ is 4 for a de Vaucouleurs' $R^{1/4}$ profile
\citep[for example,][]{devauc79}
that is often used to describe elliptical
galaxy stellar surface brightness profies, and $n$ is 1 for an exponential disk.
Values of \paramn\ for our sample range between 0.2 and 1, with only
two values larger than 1 (specifically, 1.29 for NGC 3738 and 1.65 for NGC 1569).
The family of Sersic profiles demonstrated by LITTLE THINGS \dirr\ galaxies
is shown in Figure \ref{fig-sersic}, the \HI\ profiles with the Sersic fits superposed are shown in
Figure \ref{fig-sersicprof}, and the fit parameters for each galaxy are given in Table \ref{tab-param}.

We have compared the central measured \HI\ surface mass density to \paramhi\ from the Sersic fit
relative to the uncertainties in the two quantities.
The uncertainty for the central point in the observed surface density profile was determined
from the channel rms given by \citet{hunter12}, the typical FWHM of the line profile in channels, and the area 
of the moment 0 map that was integrated. The uncertainty in the observed surface mass density
was combined in quadrature with that for \paramhi\ from the Sersic fit.
Only two galaxies, M81dwA and IC 1613, have a difference in the central mass densities that is greater than
3$\sigma$. One can see why these galaxies stand out: they have depressions in their centers.

We tried other representations in addition to the Sersic profile. In particular, we used a power law plus exponential
\begin{equation}
\Sigma_{\rm HI}(R) = \frac{S}{R^A} \exp(-R^n)
\end{equation}
where parameters of the fit are $S$, $A$, and $n$,
and a disk galaxy fit from \citet{wang14}
\begin{equation}
\Sigma_{\rm HI}(R) = \frac{I_1\exp(-R/R_s)}{1+I_2\exp(-R/R_c)}
\end{equation}
where fitting parameters are $I_1$, $I_2$, $R_s$, and $R_c$. The power law plus
exponential did not fit the LITTLE THINGS data well. The \citet{wang14} fit was
reasonable for 32 of the 40 galaxies, but generally the Sersic fit worked best
overall and that is what we use here.

\clearpage

\hspace{-0.85truein} \vspace{-2.truein}
\includegraphics[scale=.9]{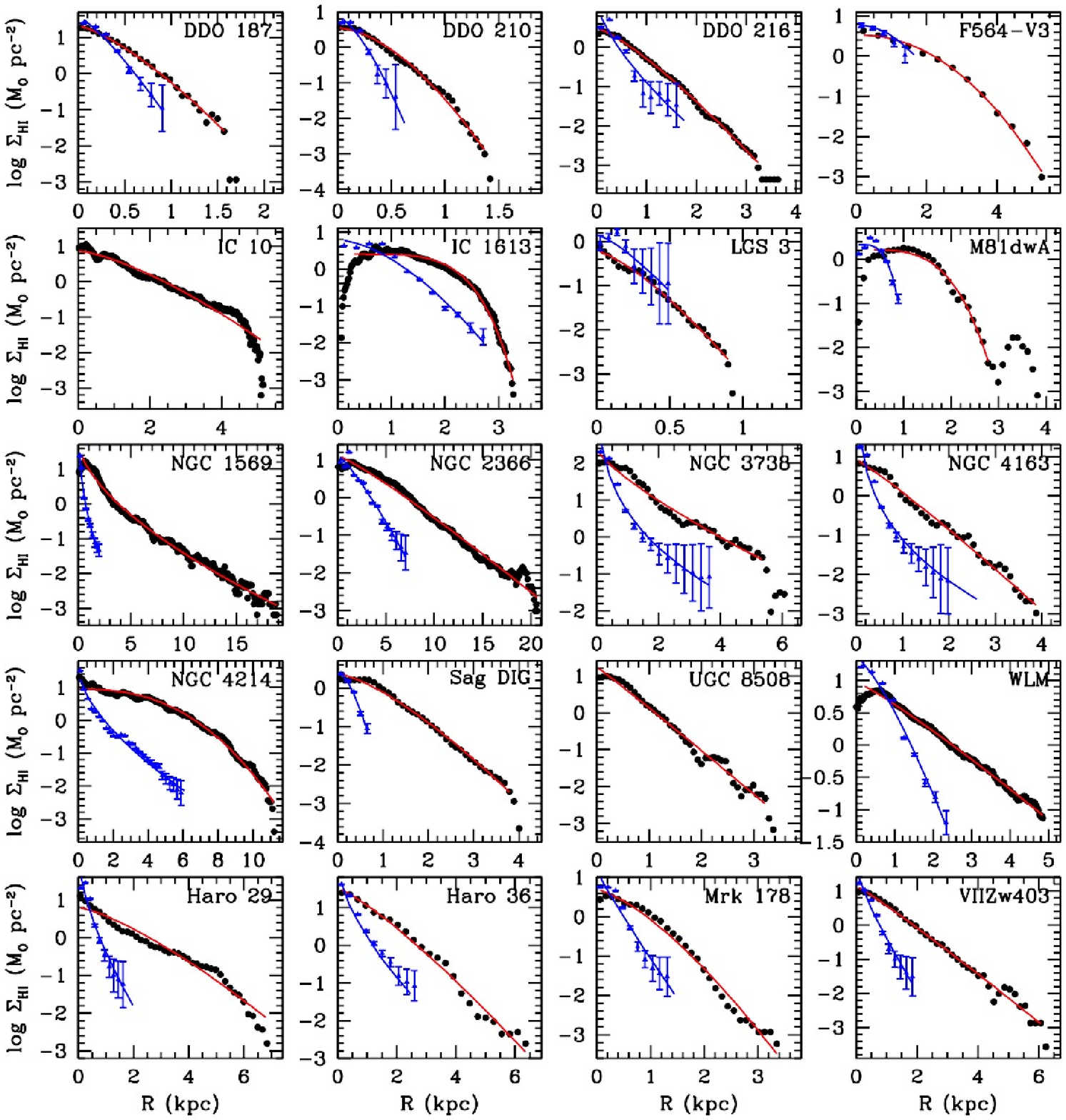}

\clearpage

\begin{deluxetable}{lccccc}
\tabletypesize{\scriptsize} \tablecaption{Sersic Fit Parameters and Radii for \HI\
\label{tab-param}}
\tablewidth{0pt}
\tablehead{
\colhead{} & \colhead{\paramhi} & \colhead{\paramr} & \colhead{}
& \colhead{\rfifty\tablenotemark{a}}
& \colhead{}  \\
\colhead{Galaxy}             & \colhead{(\unithi)}   & \colhead{(kpc)}      &  \colhead{\paramn}
& \colhead{(kpc)} & \colhead{\rrat\tablenotemark{a}} \\
}
\startdata
CVnIdwA &    1.13$\pm$0.06    &    0.99$\pm$0.06   & 0.56$\pm$0.03 & 0.77$\pm$0.02 & 0.50$\pm$0.05  \\
DDO 43   &     0.78$\pm$0.07   &     2.44$\pm$0.24  & 0.75$\pm$0.04 & 2.38$\pm$0.08 & 0.42$\pm$0.06  \\
DDO 46    &    0.95$\pm$0.04   &     2.63$\pm$0.10  & 0.42$\pm$0.02 & 1.83$\pm$0.06 & 0.52$\pm$0.05  \\
DDO 47   &     0.35$\pm$0.04   &     7.71$\pm$0.29  & 0.38$\pm$0.02 & 5.03$\pm$0.12 & 0.52$\pm$0.05 \\
DDO 50   &     0.86$\pm$0.02   &     5.67$\pm$0.09  & 0.35$\pm$0.01 & 4.12$\pm$0.12 & 0.58$\pm$0.04  \\
DDO 52    &     0.57$\pm$0.04   &    5.03$\pm$0.14  & 0.30$\pm$0.01 & 3.67$\pm$0.07 & 0.63$\pm$0.04 \\
DDO 53    &     0.86$\pm$0.05   &    1.36$\pm$0.08  & 0.65$\pm$0.03 & 1.35$\pm$0.04 & 0.52$\pm$0.05  \\
DDO 63    &    1.06$\pm$0.07   &     2.07$\pm$0.13  & 0.50$\pm$0.03 & 1.65$\pm$0.05 & 0.52$\pm$0.04  \\
DDO 69    &    0.75$\pm$0.03   &     0.58$\pm$0.02  & 0.62$\pm$0.02 & 0.56$\pm$0.02 & 0.54$\pm$0.05  \\
DDO 70    &     0.49$\pm$0.02   &    2.20$\pm$0.05  & 0.36$\pm$0.01 & 1.54$\pm$0.05 & 0.56$\pm$0.04  \\
DDO 75    &     1.23$\pm$0.07   &    1.10$\pm$0.08  & 0.68$\pm$0.03 & 1.20$\pm$0.03 & 0.53$\pm$0.06 \\
DDO 87    &     0.33$\pm$0.02   &    6.26$\pm$0.14  & 0.35$\pm$0.01 & 4.65$\pm$0.12 & 0.62$\pm$0.05  \\
DDO 101  &     0.39$\pm$0.03  &     1.76$\pm$0.05  &  0.31$\pm$0.01 & 1.20$\pm$0.03 & 0.63$\pm$0.05  \\
DDO 126  &     0.80$\pm$0.04  &     2.59$\pm$0.10  &  0.50$\pm$0.02 & 2.32$\pm$0.06 & 0.58$\pm$0.04  \\
DDO 133  &     0.76$\pm$0.05  &     2.44$\pm$0.09  &  0.37$\pm$0.02 & 1.92$\pm$0.05 & 0.63$\pm$0.04  \\
DDO 154  &     0.74$\pm$0.03  &     3.54$\pm$0.13  & 0.73$\pm$0.02 & 3.64$\pm$0.12 & 0.46$\pm$0.06  \\
DDO 155  &     1.06$\pm$0.17  &     0.31$\pm$0.08  & 1.02$\pm$0.10 & 0.57$\pm$0.01 & 0.52$\pm$0.05  \\
DDO 165   &    0.50$\pm$0.04  &     3.52$\pm$0.13  & 0.49$\pm$0.02 & 2.68$\pm$0.06 & 0.54$\pm$0.06 \\
DDO 167   &    0.53$\pm$0.05   &    1.21$\pm$0.04  & 0.27$\pm$0.02 & 0.74$\pm$0.02 & 0.60$\pm$0.04  \\
DDO 168   &    1.25$\pm$0.05   &    1.91$\pm$0.13  & 0.79$\pm$0.03 & 2.09$\pm$0.08 & 0.41$\pm$0.05  \\
DDO 187   &    1.31$\pm$0.05   &    0.39$\pm$0.03  & 0.73$\pm$0.03 & 0.43$\pm$0.01 & 0.48$\pm$0.05  \\
DDO 210   &    0.55$\pm$0.04   &    0.41$\pm$0.02  &  0.59$\pm$0.02 & 0.34$\pm$0.01 & 0.46$\pm$0.05  \\
DDO 216   &    0.49$\pm$0.03   &    0.63$\pm$0.03  &  0.80$\pm$0.02 & 0.73$\pm$0.03 & 0.46$\pm$0.05  \\
F564-V3    &    0.54$\pm$0.05    &     2.01$\pm$0.11  & 0.47$\pm$0.03 & 1.54$\pm$0.06 & 0.54$\pm$0.05  \\
IC 10          &   0.86$\pm$0.03    &     1.47$\pm$0.08  & 0.71$\pm$0.03 & 1.48$\pm$0.06 & 0.44$\pm$0.05  \\
IC 1613      &    0.40$\pm$0.01   &    2.18$\pm$0.02  &  0.20$\pm$0.00 & 1.36$\pm$0.03 & 0.62$\pm$0.04  \\
LGS 3         &  -0.21$\pm$0.04  &    0.25$\pm$0.02  &  0.74$\pm$0.04 & 0.29$\pm$0.01 & 0.52$\pm$0.05  \\
M81dwA    &    0.22$\pm$0.04   &    1.75$\pm$0.05  &  0.26$\pm$0.02 & 1.22$\pm$0.03 & 0.67$\pm$0.05  \\
NGC 1569  &   1.79$\pm$0.09   &    0.37$\pm$0.06  & 1.65$\pm$0.06 & 1.92$\pm$0.08 & 0.30$\pm$0.06  \\
NGC 2366  &    1.12$\pm$0.04  &    3.04$\pm$0.17  &  0.89$\pm$0.02 & 4.43$\pm$0.14 & 0.51$\pm$0.06  \\
NGC 3738  &    2.39$\pm$0.12   &   0.44$\pm$0.12  &  1.29$\pm$0.12 & 1.09$\pm$0.04 & 0.36$\pm$0.06  \\
NGC 4163  &    0.92$\pm$0.08   &   0.56$\pm$0.07  &  0.90$\pm$0.05 & 0.68$\pm$0.02 & 0.37$\pm$0.05  \\
NGC 4214  &    0.95$\pm$0.02   &   4.89$\pm$0.11  &  0.40$\pm$0.01 & 3.74$\pm$0.10 & 0.58$\pm$0.05 \\
Sag DIG      &   0.37$\pm$0.04   &   0.92$\pm$0.05  &  0.72$\pm$0.03 & 0.95$\pm$0.03 & 0.48$\pm$0.05  \\
UGC 8508  &    1.24$\pm$0.09   &   0.40$\pm$0.06  &  0.97$\pm$0.06 & 0.62$\pm$0.02 & 0.46$\pm$0.05 \\
WLM            &   0.97$\pm$0.02   &   1.22$\pm$0.04  &  0.88$\pm$0.02 & 1.62$\pm$0.06 & 0.47$\pm$0.05  \\
Haro 29      &    0.80$\pm$0.09   &    1.64$\pm$0.22  &  0.75$\pm$0.07 & 1.84$\pm$0.08 & 0.43$\pm$0.04  \\
Haro 36      &    1.51$\pm$0.11   &    0.97$\pm$0.14  &  0.82$\pm$0.06 & 1.32$\pm$0.05 & 0.49$\pm$0.06  \\
Mrk 178      &    0.68$\pm$0.09   &    0.68$\pm$0.07  &  0.71$\pm$0.05 & 0.79$\pm$0.02 & 0.55$\pm$0.05 \\
VIIZw403   &     1.15$\pm$0.08   &    0.76$\pm$0.09  &  0.93$\pm$0.05 & 1.10$\pm$0.04 & 0.44$\pm$0.05  \\
\enddata
\tablenotetext{a}{\rfifty\ is the radius that contains 50\% of the \HI\ of the galaxy, and \rninety\ contains 90\%.
The uncertainties in these radii are determined from finding the radius at the indicated \HI\ mass plus and minus its 
uncertainty. The uncertainty in the \HI\ mass depends
on the channel rms given by \citet{hunter12}, the typical FWHM of the line profile in channels, and the area integrated.}
\end{deluxetable}

\subsection{FUV Luminosity}

We have examined the SFR interior to \rbr\ compared to that exterior to the break.
We used FUV as a tracer of the SFR because dust absorption is usually small in
dIrr galaxies. We normalized the interior and exterior FUV luminosities in two
ways: 1) relative to the area over which the FUV has been integrated, and 2)
relative to the $V$-band luminosity in the same region. Normalizing to the $V$-band
luminosity is comparable to normalizing the flux from young stars to that from older
stellar populations. We used the {\it GALEX} FUV and $V$-band surface photometry for
the LITTLE THINGS dwarfs. The effective wavelength of the FUV filter was 1516 \AA\
and the resolution was 4.0\arcsec. The FUV ratios are given in Table \ref{tab-rats}.

In order to compare this measurement of the dwarfs with those of spiral galaxies also with
breaks in their surface brightness profiles, we used
galaxies that are part of the {\it GALEX} Nearby Galaxy Survey (NGS) \citep{fuvspirals},
which has obtained {\it GALEX} images of a wide range of spirals.
Using the Lowell Observatory Hall 1.1 m telescope in January 2014, we obtained deep $V$-band images 
of 8 of the spirals in the NGS list. Galaxies were chosen to be not too edge on,
observable from Flagstaff in January, and small enough for the field of view of the detector we used (19\arcmin$\times$19\arcmin).
In addition the galaxies were selected to cover the range of morphological type Sa, Sb, Sc, and Sd.
Ultimately three of these galaxies were discarded, two because they did not have FUV images and one
because it did not show a break in the $V$-band surface brightness profile. 
This left a sample of 5: one each of Sb, Sc, and Sd classifications, and two Sa galaxies.
A sample of 5 is too small for characterizing either the mean or the dispersion in spiral galaxies of the SFR ratio parameter
discussed here, especially as a function of spiral type, but we include these observations as a hint at
the properties of spiral galaxies compared to those of the \dirr\ galaxies.
In the Appendix we present the FUV and $V$ surface photometry 
and their derived properties of these 5 spirals.

\begin{deluxetable}{lccc}
\tabletypesize{\scriptsize}
\tablecaption{FUV interior and exterior to \rbr\ \label{tab-rats}}
\tablewidth{0pt}
\tablecolumns{4}
\tablehead{
\colhead{}
& \colhead{}
& \multicolumn{2}{c}{log Interior/Exterior\tablenotemark{b}} \\
\colhead{Galaxy}
& \colhead{Break Type\tablenotemark{a}}
& \colhead{FUV/Area}
& \colhead{FUV/$V$}  \\
}
\startdata
CVnIdwA     & FI            &  $1.05\pm0.19$  &  $0.42\pm0.07$ \\
DDO 43      & II              &  $0.89\pm0.03$  & $-0.19\pm$0.05 \\
DDO 46      & II              &  $1.02\pm0.04$  & $-0.11\pm0.03$ \\
DDO 47      &  I              & \nodata & \nodata \\
DDO 50      & II              &  $0.84\pm0.02$  & $-0.21\pm0.01$ \\
DDO 52      & II              &  $0.61\pm0.07$  & $-0.06\pm0.11$ \\
DDO 53      & FI             &  $1.41\pm0.08$  &  $0.54\pm0.01$ \\
DDO 63      & FI             &  $0.69\pm0.01$  &  $0.13\pm0.02$ \\
DDO 69      & FI             &  $0.77\pm0.03$  &  $0.08\pm0.01$ \\
DDO 70      & FI             &  $1.03\pm0.07$  &  $0.21\pm0.01$ \\
DDO 75      & FI             &  $0.56\pm0.01$  & $-0.02\pm0.01$ \\
DDO 87      & II              &  $0.47\pm0.11$  & $-0.27\pm0.02$ \\
DDO 101     & II             &  $0.74\pm0.04$  & $ 0.04\pm0.03$ \\
DDO 126     & FI+II        &  $0.66\pm0.05$  & $-0.04\pm0.01$ \\
DDO 133     & II             &  $1.11\pm0.04$  &  $0.06\pm0.04$ \\
DDO 154     & II             &  $0.68\pm0.04$  & $-0.05\pm0.01$ \\
DDO 155     & FI            & \nodata & \nodata \\
DDO 165     & II             & \nodata & \nodata \\
DDO 167     & II             &  $1.08\pm0.08$  &  $0.28\pm0.04$ \\
DDO 168     & FI            &  $0.91\pm0.01$  &  $0.16\pm0.01$ \\
DDO 187     & II             &  $1.06\pm0.03$  &  $0.20\pm0.02$ \\
DDO 210     & I              & \nodata & \nodata \\
DDO 216     & II             &  $1.69\pm0.92$  &  $0.69\pm0.92$ \\
F564-V3     & II              &  $0.39\pm0.06$  & $-0.16\pm0.05$ \\
IC 10          & III             & \nodata & \nodata \\
IC 1613     & II               &  $0.87\pm0.02$  &  $0.01\pm0.00$ \\
LGS 3       & II               &  $1.20\pm0.48$  &  $0.75\pm0.48$ \\
M81dwA      & FI           &  $0.50\pm0.02$  & $-0.04\pm0.06$ \\
NGC 1569    & III          &  $1.68\pm0.03$  &  $0.42\pm0.02$ \\
NGC 2366    & II           &  $1.21\pm0.01$  &  $0.15\pm0.00$ \\
NGC 3738    & III          &  $2.15\pm0.02$  &  $0.87\pm0.02$ \\
NGC 4163    & III          &  $2.19\pm0.04$  &  $0.67\pm0.02$ \\
NGC 4214    & III          &  $1.62\pm0.01$  &  $0.05\pm0.00$ \\
SagDIG        & II           &  $0.92\pm0.13$  &  $0.22\pm0.10$ \\
UGC 8508    & II           & \nodata & \nodata \\
WLM           & II             &  $1.10\pm0.03$  &  $0.32\pm0.00$ \\
Haro 29     & III              &  $1.75\pm0.08$  &  $0.01\pm0.10$ \\
Haro 36     & II               &  $1.46\pm0.03$  &  $0.86\pm0.02$ \\
Mrk 178     & FI+III        &  $1.30\pm0.02$  &  $0.31\pm0.01$ \\
VIIZw 403   & III             &  $1.67\pm0.08$  &  $0.76\pm0.08$ \\
\enddata
\tablenotetext{a}{Type of surface brightness profile break in the $V$-band \citep{herrmann13}. ''FI'' refers to a profile in which
there is a short central segment that is flat or rising and then the profile drops off exponentially.
''II'' refers to a downward break, and ''III'' to an upward bend. ''FI+II'' or ''FI+III'' refers to a profile with two breaks.
''I'' means that the profile does not show any break. Galaxies without breaks and those without FUV images do
not have ratios of FUV interior to exterior of the break.}
\tablenotetext{b}{The FUV flux is normalized by the area over which it is measured, ``FUV/Area,'' or by
the $V$-band flux measured over the same area, ``FUV/V.'' The ratio that is given is FUV/Area or FUV/$V$
measured interior to the surface brightness profile break to that measured exterior to the break.}
\end{deluxetable}

\subsection{FUV Intensity Profiles}
\label{sersic-fuv}

The FUV radial profiles were also fit with Sersic functions. The profile
measurements and fitted curves are shown in Figure \ref{fig-sersicprof} in blue and
the parameter results are in Table \ref{tab-paramfuv}. 
The Sersic fit parameters are \paramfuv, \paramrfuv, and \paramnfuv.

\begin{deluxetable}{lccc}
\tabletypesize{\scriptsize} \tablecaption{Sersic Fit Parameters for FUV Profiles
\label{tab-paramfuv}}
\tablewidth{0pt} \tablehead{ \colhead{Galaxy} & \colhead{\paramfuv} &
\colhead{\paramrfuv} &
\colhead{\paramnfuv}\\
\colhead{} & \colhead{(mag arcsec$^{-2}$)}& \colhead{(kpc)}      & \colhead{}
\\
} \startdata
CVnIdwA  & $24.96\pm0.37$  & $0.55\pm0.03$  & $0.52\pm0.09$ \\
DDO 43   & $24.93\pm0.05$  & $2.91\pm0.03$  & $0.47\pm0.01$ \\
DDO 46   & $24.65\pm0.27$  & $0.95\pm0.05$  & $0.71\pm0.09$ \\
DDO 47   & $25.36\pm0.14$  & $1.72\pm0.07$  & $0.85\pm0.12$ \\
DDO 50   & $24.01\pm0.07$  & $2.29\pm0.04$  & $0.57\pm0.06$ \\
DDO 52   & $25.94\pm0.16$  & $2.20\pm0.12$  & $0.55\pm0.14$ \\
DDO 53   & $23.31\pm0.33$  & $0.28\pm0.01$  & $1.06\pm0.07$ \\
DDO 63   & $25.50\pm0.14$  & $1.61\pm0.05$  & $0.43\pm0.09$ \\
DDO 69   & $25.88\pm0.12$  & $0.44\pm0.01$  & $0.45\pm0.05$ \\
DDO 70   & $23.73\pm0.13$  & $0.28\pm0.01$  & $1.02\pm0.06$ \\
DDO 75   & $24.61\pm0.11$  & $0.98\pm0.02$  & $0.21\pm0.07$ \\
DDO 87   & $26.79\pm0.15$  & $2.72\pm0.11$  & $0.50\pm0.12$ \\
DDO 101  & $26.70\pm0.23$  & $1.40\pm0.05$  & $0.34\pm0.06$ \\
DDO 126  & $25.35\pm0.13$  & $1.60\pm0.05$  & $0.55\pm0.08$ \\
DDO 133  & $25.92\pm0.14$  & $1.81\pm0.03$  & $0.41\pm0.05$ \\
DDO 154  & $24.72\pm0.07$  & $1.13\pm0.02$  & $0.50\pm0.04$ \\
DDO 167  & $24.08\pm0.37$  & $0.27\pm0.02$  & $0.64\pm0.07$ \\
DDO 168  & $24.57\pm0.10$  & $1.23\pm0.02$  & $0.62\pm0.04$ \\
DDO 187  & $24.31\pm0.16$  & $0.22\pm0.01$  & $0.79\pm0.04$ \\
DDO 210  & $25.82\pm0.24$  & $0.17\pm0.01$  & $0.67\pm0.05$ \\
DDO 216  & $25.30\pm0.48$  & $0.10\pm0.01$  & $1.47\pm0.19$ \\
F564-V3  & $26.79\pm0.01$  & $1.27\pm0.01$  & $0.47\pm0.01$ \\
IC 1613  & $24.76\pm0.17$  & $0.89\pm0.02$  & $0.60\pm0.05$ \\
LGS 3    & $28.84\pm0.25$  & $0.22\pm0.02$  & $0.76\pm0.27$ \\
M81dwA   & $26.16\pm0.16$  & $0.67\pm0.02$  & $0.26\pm0.07$ \\
NGC 1569  & $14.22\pm0.41$  & $0.0019\pm0.0002$  & $2.94\pm0.13$ \\
NGC 2366  & $23.71\pm0.17$  & $1.59\pm0.05$  & $0.81\pm0.07$ \\
NGC 3738  & $15.47\pm0.59$  & $0.0008\pm0.0001$  & $3.23\pm0.15$ \\
NGC 4163  & $14.42\pm0.64$  & $0.00\pm0.00$  & $4.17\pm0.25$ \\
NGC 4214  & $20.41\pm0.16$  & $0.19\pm0.01$  & $1.61\pm0.09$ \\
Sag DIG   & $24.42\pm0.11$  & $0.41\pm0.01$  & $0.62\pm0.04$ \\
WLM       & $24.35\pm0.15$  & $0.61\pm0.02$  & $0.75\pm0.04$ \\
Haro 29   & $20.43\pm0.40$  & $0.07\pm0.01$  & $1.52\pm0.12$ \\
Haro 36   & $20.37\pm0.18$  & $0.15\pm0.01$  & $1.56\pm0.06$ \\
Mrk 178   & $22.31\pm0.32$  & $0.17\pm0.01$  & $1.08\pm0.09$ \\
VIIZw403  & $21.49\pm0.29$  & $0.09\pm0.01$  & $1.25\pm0.08$ \\
\enddata
\label{sersicFUVtable}
\end{deluxetable}


\subsection{\HI\ Rotation Curves}


\citet{oh15} have determined the rotation curves of a subset of the LITTLE THINGS
galaxies. We include 17 of those galaxies in this analysis. The remaining rotation
curves were excluded for one or several of the following reasons: 1) visually, they
did not have a horizontal velocity asymptote, so that although a fit was possible,
at most only lower bounds
for 
several parameters could be obtained from the data, 2) they had a profile with 3
distinct sections, which could not be fit well, or 3) they were concave with no flat
section. For the galaxies included here, we fit the rotation curve with the
following function, motivated by \citet{courteau97}:
\begin{equation}
V(R)=V_c\frac {(1+\frac{R_t}{R})^\beta} {(1+(\frac{R_t}{R})^\gamma)^{1/\gamma}}
\label{eq-rotcurv}
\end{equation}
The fit parameters are $V_c$, the asymptotic velocity at large radii; $R_t$, the
transition radius at which the increase in rotation speed with radius ends;
$\gamma$, which regulates the sharpness of the transition; and $\beta$, which allows
for a downward slope in the outer galaxy. We fit both with $\beta$ as a free
parameter and with fixed $\beta=0$. The fits were similar, so we elected to work
with the $\beta=0$ fits. We assume that $V(0)=0$. The python function {\sc
scipy.optimize.curve\_fit} was used to fit the data, and the uncertainties in the
fitting parameters were  given by the diagonal components of the covariance matrix.

\begin{figure}[t!]
\epsscale{0.9}
\plotone{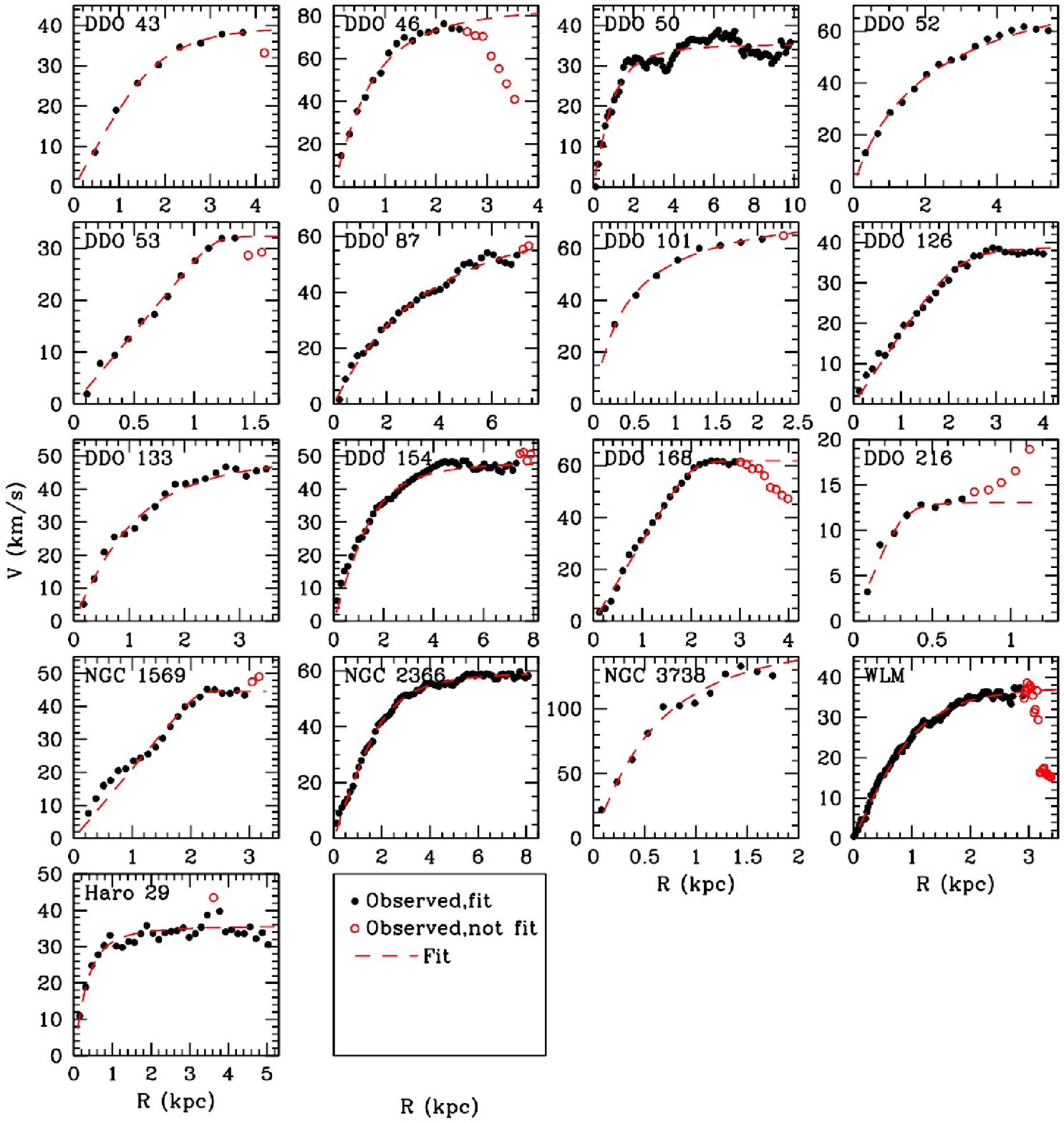}
\caption{Rotation curves from \citet{oh15} with fits from Equation \ref{eq-rotcurv} with $\beta=0$ superposed as
dashed lines. Open red points were omitted from the fit.
\label{fig-rotcurv}}
\end{figure}

The rotation curve fit parameters are given in Table \ref{tab-rotparam} and their
fits are shown in Figure \ref{fig-rotcurv}. In some cases a data point or two,
usually at the end, in the rotation curve were inconsistent with the rest of the
points and they were eliminated from the fit. Additionally, DDO 46 and DDO 168
showed asymptotically flat behavior with a concave section at large radii. Although
$\beta$ allows for some concavity in the rotation curve, we were unable to fit the
entire rotation curve and so fit only the portion of the rotation curve interior to
the concave section. Similarly, DDO 216's rotation curve plateaus and then begins
rising again at larger radii, and the rising outer part of the curve is not included
in the fit.

\begin{deluxetable}{lccc}
\tabletypesize{\scriptsize} \tablecaption{Rotation Curve Fit Parameters
\label{tab-rotparam}}
\tablewidth{0pt} \tablehead{
\colhead{Galaxy}        & \colhead{\paramrt (kpc)} & \colhead{\paramvc (\kms)} & \colhead{\paramgamma} \\ 
} \startdata
DDO 43     &  2.01$\pm$0.07  &  39.9$\pm$1.3     &   3.15$\pm$0.65 \\
DDO 46     &  0.95$\pm$0.05  &  85.9$\pm$4.8     &   1.63$\pm$0.29 \\
DDO 50     &  1.30$\pm$0.09  &  35.5$\pm$0.5     &   2.00$\pm$0.25 \\
DDO 52     & 2.00$\pm$0.10   &  91.7$\pm$10.6   &    0.89$\pm$0.13 \\
DDO 53     & 1.17$\pm$0.02   &  32.3$\pm$0.6     &  17.65$\pm$8.64 \\
DDO 87     & 4.18$\pm$0.34   &  76.8$\pm$13.0   &    1.22$\pm$0.34 \\
DDO 101   & 0.38$\pm$0.05   &  78.5$\pm$6.3     &    0.94$\pm$0.18 \\
DDO 126   & 2.23$\pm$0.05   &  38.8$\pm$0.8     &    5.72$\pm$1.15 \\
DDO 133   & 1.38$\pm$0.10   &  52.0$\pm$3.3     &   1.69$\pm$0.38 \\
DDO 154   & 1.93$\pm$0.10   &  48.3$\pm$0.6     &   2.30$\pm$0.27 \\
DDO 168   & 2.02$\pm$0.02   &  61.9$\pm$0.5     & 10.97$\pm$1.84 \\
DDO 216   & 0.32$\pm$0.04   &  13.1$\pm$1.3     &   5.07$\pm$4.81 \\
NGC 1569 &  2.11$\pm$0.05  &  44.6$\pm$0.5     & 20.88$\pm$15.38 \\
NGC 2366 &  2.38$\pm$0.03  &  59.5$\pm$0.3     &   2.69$\pm$0.12 \\
NGC 3738 &  0.76$\pm$0.17  & 160.9$\pm$37.3  &   1.43$\pm$0.91 \\
WLM          & 1.27$\pm$0.03    &  38.0$\pm$0.5    &   2.55$\pm$0.23 \\
Haro 29     & 0.47$\pm$0.06   &  35.8$\pm$0.9     &    1.76$\pm$0.39 \\
\enddata
\end{deluxetable}

\clearpage

\subsection{Summary of parameters}

In Section \ref{sec-profiles} we have described the parameters that we use to characterize the
stellar, gas, and star formation surface density profiles and related parameters
for purposes of examining the relationships between these galactic components.
The parameters are listed and described as a reference to the reader in Table \ref{tab-summary}.

\begin{deluxetable}{lllll}
\tabletypesize{\scriptsize} \tablecaption{Summary of disk parameters
\label{tab-summary}}
\rotate
\tablewidth{0pt} \tablehead{
\colhead{Disk component} & \colhead{Quantity} & \colhead{Description} & \colhead{Note} & \colhead{Ref} \\
} 
\startdata
\HI                   & \paramhi  & Extrapolated central surface mass density & Sersic fit                                       & Table \ref{tab-param} \\
                        & \paramr   & Characteristic radius                           & Sersic fit                                                                & Table \ref{tab-param} \\
                        & \paramn  & Curvature of profile                              & Sersic fit                                                                & Table \ref{tab-param} \\
                        & \rrat        & Central concentration of \HI                  & Low ratio for more central concentration               & Table \ref{tab-param} \\
Stars                & \rd           & Disk scale length                                 &                                                                                & \citet{herrmann13}  \\
                        & \rbr          & Break radius                                       & Where the surface brightness profile changes slope                             & \citet{herrmann13} \\
                        & $C_{31}$ & Central concentration of stellar mass & Larger $C_{31}$, more centrally concentrated      & \citet{zhang12} \\
FUV                 & \paramfuv  & Extrapolated central surface brightness  & Sersic fit                                                           & Table \ref{tab-paramfuv} \\
                        & \paramrfuv   & Characteristic radius                           & Sersic fit                                                                  & Table \ref{tab-paramfuv} \\
                        & \paramnfuv  & Curvature of profile                             & Sersic fit                                                                   & Table \ref{tab-paramfuv} \\ 
                        & FUV/$V$ before/after & Change in SF activity at \rbr & Ratio of normalized FUV before and after \rbr   & Table \ref{tab-rats} \\
                        & \rfuvknot  & Radial extent of furthest FUV knot    &                                                                                   & \citet{outerfuv} \\
                        & \logfuvrat & Concentration of SF activity              & Ratio of FUV in 1\rd\ to that in 1-3\rd                        & Table \ref{tab-gal} \\
                        & \sfr           & Integrated SFR                                 & Measured in FUV, normalized to 1\rd\ area                       & \citet{hunter12} \\
\ha                   & \rha          & Radial extent of \ha                           &                                                                                  & \citet{he04} \\
Rotation curve & \paramvc & Asymptotic velocity                            &                                                                                  & Table \ref{tab-rotparam} \\
                        & \paramrt   & Transition radius                               & Where rotation speed levels off or increases slowly & Table \ref{tab-rotparam} \\
                        & \paramgamma & Sharpness of transition             &                                                                                  & Table \ref{tab-rotparam} \\
\enddata
\end{deluxetable}

\section{Results} \label{sec-results}

\subsection{Comparisons with \HI\ surface density profiles}

\subsubsection{$M_V$}

\begin{figure}[t!]
\epsscale{0.8}
\plotone{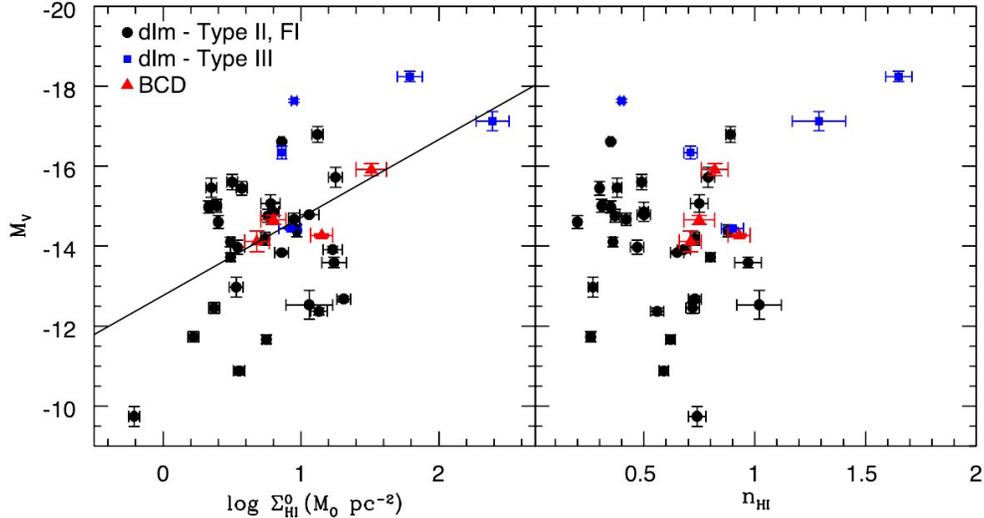}
\caption{Integrated galactic $M_V$ plotted against Sersic parameters \paramhi\ and \paramn\ that describe the \HI\ surface density profile.
There is a trend $M_V = (-12.77\pm0.52) -(1.95\pm0.55)\log \Sigma_{\rm HI}^0$, 
correlation coefficient 0.5, standard deviation 1.5.
The trend is in the sense that more luminous
dwarfs have higher central atomic gas densities. There is no trend of $M_V$ with \paramn\
correlation coefficient of 0.2). 
\label{fig-mv}}
\end{figure}

In Figure \ref{fig-mv} we plot integrated galactic $M_V$ against key Sersic parameters: \paramhi\
and \paramn. We see that there is a trend of $M_V$ with \paramhi, with a correlation coefficient of 0.5. 
The relationship ($M_V = (-12.77\pm0.52) -(1.95\pm0.55)\log \Sigma_{\rm HI}^0$) is one in which more luminous
dwarfs tend to have higher central atomic gas densities.
There is no trend of $M_V$ with \paramn, the parameter that describes how the gas falls off with radius.

\subsubsection{Characteristic radii}

\begin{figure}[t!]
\epsscale{0.9}
\plotone{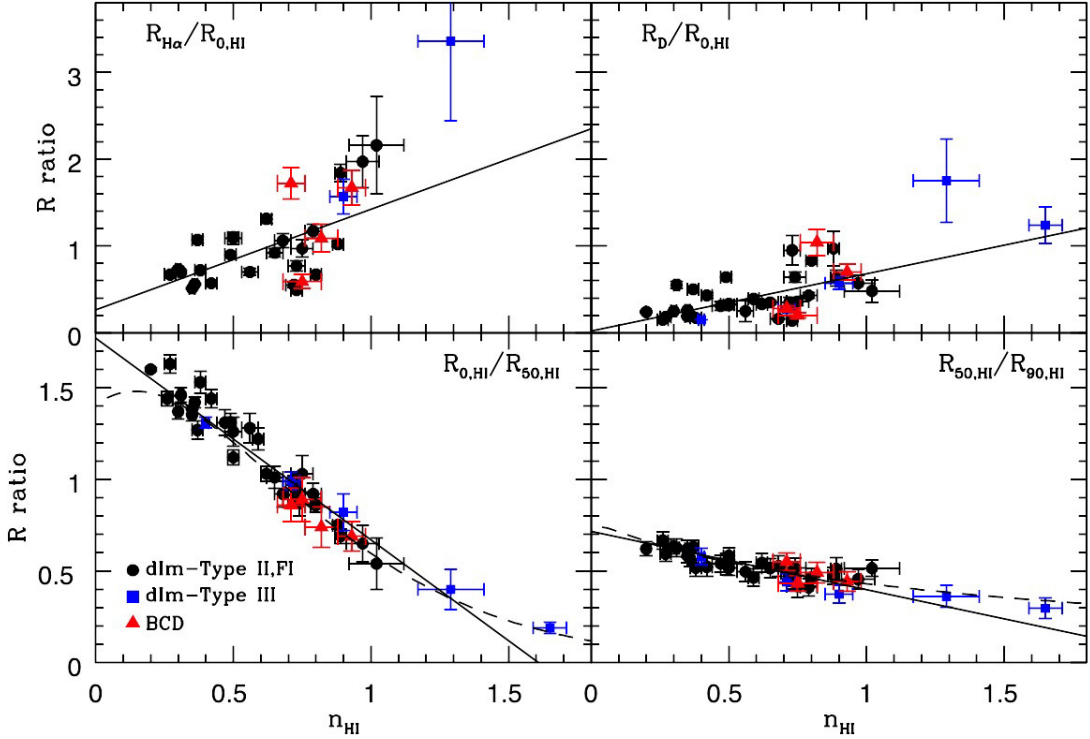}
\vskip -0.2truein
\caption{Plots of ratios of characteristic radii (see Table \ref{tab-param})
against Sersic \paramn. \paramr\ is the normalizing radius in the \HI\ Sersic profile.
The two most discrepant galaxies (DDO 155 and NGC 3738) are not included in the fits
shown in the upper two panels.
{\it Upper left:} \rha\ is the radius at which the \HII\ region furthest from the center of the galaxy is located.
The solid line is a fit to the data:
$R_{H\alpha}/R_{0,HI} = (0.26\pm0.20) + (1.16\pm0.30)n_{HI}$, and the correlation coefficient is 0.6
and the standard deviation is 0.4.
Galaxies with more distant \HII\ regions relative to \paramr\ have higher \paramn.
{\it Upper right:} \rd\ is the stellar disk scale length measured in $V$ (Herrmann \et\ 2013).
The solid line is a fit to the data:
$R_D/R_{0,HI} = (0.022\pm0.082) + (0.66\pm0.12)n_{HI}$, and the correlation coefficient is 0.7
and the standard deviation is 0.2.
Bigger disk scale lengths relative to \paramr\ are associated with higher \paramn\ \HI\ profiles.
{\it Lower left:} \rfifty\ is the radius that contains half of the \HI\ gas.
The solid line is a fit to the data (correlation coefficient of 0.97, standard deviation of 0.09):
$R_{0,HI}/R_{50,HI} = (1.77\pm0.03) - (1.10\pm0.05)n_{HI}$.
The smaller \paramr\ is relative to \rfifty, the bigger \paramn\ is.
{\it Lower right:} \rninety\ is the radius that contains 90\% of the \HI\ gas.
The solid line is a fit to the data (correlation coefficient of 0.9, standard deviation of 0.04):
$R_{50,HI}/R_{90,HI} = (2.26\pm0.15) - (3.15\pm0.30)n_{HI}$.
The dashed curves in the lower panels are expected for a pure Sersic disk.
\label{fig-rat}}
\end{figure}

What do the parameters \paramn\ and \paramr\ actually mean in a dwarf galaxy?
In Figure \ref{fig-rat} we plot various ratios of characteristic radii against \paramn.
The radii we use include
\paramr, the characteristic radius in the \HI\ Sersic profile;
\rha, the furthest radius at which \ha\ emission is detected; \rd, the
stellar disk scale length measured in $V$; \rfifty, the radius that contains half
of the \HI\ gas; and \rninety, the radius that contains 90\% of the \HI\ gas (see
Table \ref{tab-param}). A low ratio of \rrat\ indicates that the inner 50\%
of the \HI\ is more centrally concentrated or that the outer 40\% of the \HI\ is
more extended. There is a strong relationship between all ratios and \paramn.
We see that galaxies with higher \paramn\ have, relative to \paramr, more far-flung
\HII\ regions, bigger $V$-band disk scale lengths, and larger \rfifty.

The dashed lines in the bottom panels are from integrals over the Sersic
function itself, $\int_0^R \exp(-[R/R_0]^{1/n})2\pi RdR$, and show the expected ratios of
radii for \HI\ if the \HI\ profiles are perfect Sersic fits throughout. We see that
the galaxy data generally follow the dashed curves for $R_{0}/R_{50}$ vs.\ \paramn\
and \rrat\ vs.\ \paramn, including at higher \paramn\ where the dashed
curve deviates from the linear fit to the galaxy points. This implies that the \HI\
surface density profiles are well fit by a Sersic disk. The exception is in the
panel for $R_{0}/R_{50}$ vs.\ \paramn, at low end of \paramn, where the curve turns over
and two galaxies (DDO 167 and IC 1613) do not follow that turnover. 
Both galaxies have depressions or holes in their \HI\ in their centers.
However, the gas beyond the holes is fit with a Sersic profile with very low \paramn,
which Figure \ref{fig-sersic} shows corresponds to overall more centrally concentrated \HI\ (at least outside the holes),
and the smaller $R_{50,HI}$ relative to \paramr, is consistent with 
the gas being more centrally concentrated compared to a pure Sersic disk.

The top two panels mix optical and \HI\ radii, and because the furthest
H$\alpha$ region is typically at a distance of between 2 and 3 $V$-band disk scale
lengths, the top left panel is showing about the same correlation as the top right
panel. This correlation indicates that the \HI\ scale length increases relative to the
optical scale length as the \HI\ profile becomes relatively flatter in the inner
regions. In Section \ref{sect:dark}, we explain this correlation as the result of a
conversion of \HI\ into molecules in the inner regions, which flattens the \HI\
profile more than the total gas profile, lowering \paramn\ and increasing \paramr\ at the
same time relative to \rd.

\subsubsection{FUV radial profiles}

\begin{figure}[t!]
\epsscale{0.8}
\plotone{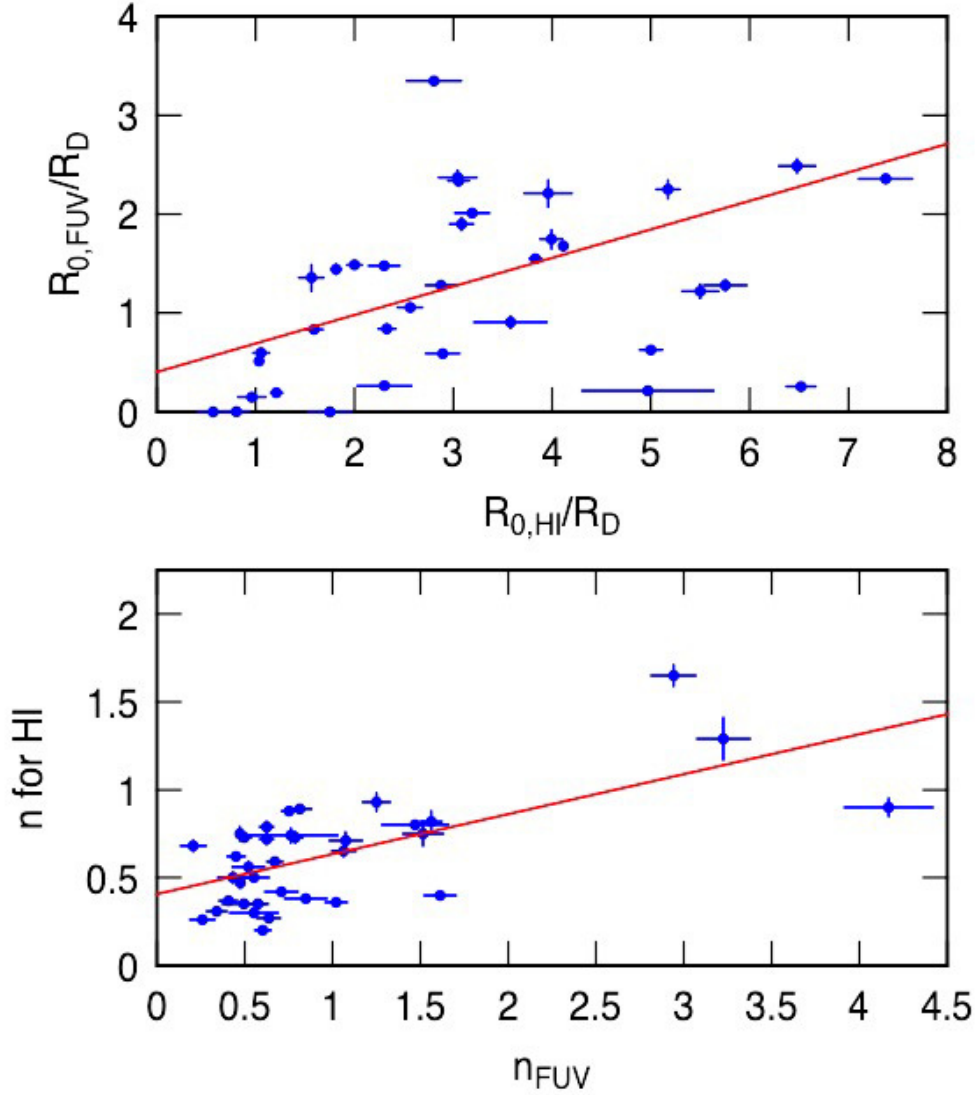}
\vskip -0.25truein
\caption{Bottom: The Sersic index for \HI\ versus the Sersic index for FUV, showing a correlation
with $n_{HI}\sim0.6 n_{\rm FUV}$. Top: Scale length comparison for FUV (\paramrfuv) and \HI\ (\paramr), relative to the $V$-band
scale length, $R_{\rm D}$. FUV disks average a factor of $\sim2$ smaller than \HI\ disks.}
\label{sersic_fuv_vs_hi}
\end{figure}

The Sersic indices for FUV,
\paramnfuv, and for HI, \paramn, and the scale lengths for FUV, \paramrfuv, and
for HI, \paramr, relative to the $V$-band scale length, \rd, are plotted in
Figure \ref{sersic_fuv_vs_hi}. Both the Sersic indices and the scale lengths
correlate with each other for the FUV and \HI, with \paramn\ about 0.6 times
\paramnfuv. The actual fits are $n_{HI}=(0.41\pm0.10)+(0.23\pm0.08)n_{\rm FUV}$ and
$R_{\rm 0,FUV}/R_{\rm D}=(0.40\pm 0.71)+(0.29\pm0.19)(R_0/R_{\rm D})$. The
correlations imply that when the \HI\ turns over in the inner region of a galaxy (low
\paramn), the FUV does also, although the FUV scale length is about half the \HI\
scale length.


\subsubsection{Integrated star formation rates}

\begin{figure}
\epsscale{0.6}
\plotone{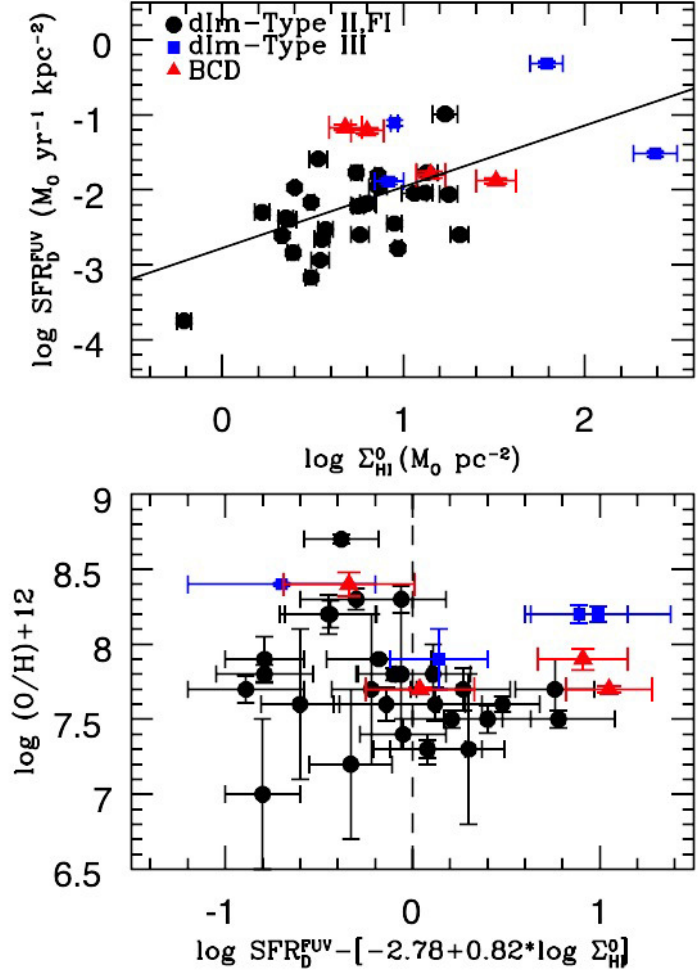}
\caption{{\it Top:} Logarithm of the integrated SFR per unit $V$-band
disk area plotted against the extrapolated central \HI\ density from
the Sersic fit.
Uncertainties are generally smaller than the point size.
The solid line is a fit to the data, with a correlation coefficient of 0.59 and standard deviation of 0.5: 
$\log {\rm SFR_D^{FUV}} = (-2.78\pm0.18) + (0.82\pm0.19) \log\Sigma_{HI}^0$.
If the three outliers are not include the slope becomes $0.3\pm0.4$ and the correlation coefficient is 0.12.
{\it Bottom:} Oxygen abundance plotted against SFR minus the fit in the upper panel.
The vertical dashed line delineates the locus with no scatter from the upper panel fit.
There appears to be no relationship between the scatter around the fit in the upper panel and metallicity.
\label{fig-sfr}}
\end{figure}

In the top panel of Figure \ref{fig-sfr} we show a correlation between the logarithm
of the total galactic FUV SFR normalized to the area inside the $V$-band scale
length, \sfr, and the logarithm of the extrapolated central \HI\ surface density, \paramhi, from the Sersic fit.
The SFR is normalized in order to compare galaxies of different sizes and
masses. We see that galaxies with higher extrapolated central \HI\ surface densities have higher
galactic SFRs per unit disk area. A least-squares fit to the data yields
the relationship
\begin{equation}
\log {\rm SFR_D^{FUV}} = (-2.78\pm0.18) + (0.82\pm0.19) \log\Sigma_{\rm HI}^0.
\end{equation}
The rms of the fit is 0.54. Thus, the galaxy-wide area-normalized SFR is proportional to the
extrapolated central \HI\ density to the 0.8 power.

Gas in dwarf galaxies appears to be dominated by the atomic phase \citep{kenney88}.
There should be molecules present, as observed in the SMC \citep{bolatto11} and in
other dwarfs such as NGC 1569 \citep{taylor99} and IC 10
\citep{ohta88,ohta92,wilson91,leroy06}, but the molecular abundances traced by
CO could be low compared to \HI\ \citep[e.g.,][]{bigiel08,rubio15}. Most dwarfs in
our survey also have a total \HI\ mass that is larger than the stellar mass
\citep{zhang12}, which is rarely true for spirals. Thus, the near-linear
relationship between integrated SFR per unit area and extrapolated central \HI\ density \paramhi\
suggests a close connection between star formation and atomic gas that is not
expected for spirals. In Section \ref{sect:dark}, we estimate the H$_2$ surface
density that is present in the central regions.

Figure \ref{fig-sfr} has considerable scatter in the distribution of points, so we
wondered if there could be a physical origin for this scatter: are the high or low
points systematically high or low because of some dependence on an additional
parameter? The most obvious additional parameter is metallicity. The oxygen
abundance given by \oh\ varies by 1.8 dex in our sample.  To examine this, we plot
in Figure \ref{fig-sfr} the oxygen abundance versus the quantity $\log {\rm
SFR_D^{FUV}}-(-2.78+0.82\log \Sigma_{\rm HI}^0)$, which is the difference between
the SFR and the average dependence on \paramhi. We see no trend. This lack of a
trend makes sense if most of the molecular regions are H$_2$ with relatively little
CO, and if the H$_2$ always has sufficient time to form even at a low relative
dust abundance. Then the relative molecular fraction would not depend much on
metallicity, and the residual gas, viewed here in \HI, would be relatively
independent of metallicity too.

\subsubsection{Concentration of stars and gas}

\begin{figure}[t!]
\epsscale{0.5}
\vskip -0.7truein
\plotone{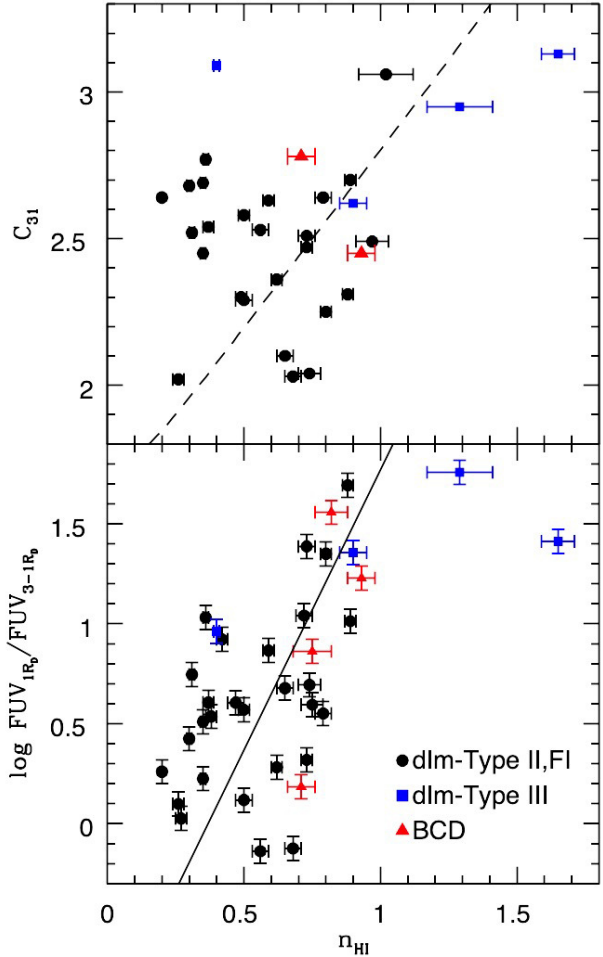}
\vskip -0.2truein
\caption{
{\it Top:}
Plot of stellar concentration index $C_{31}$ \citep{zhang12} against \paramn.
Higher $C_{31}$ means a higher central concentration of stars.
The dashed line shows what to expect if the surface densities of stars and gas have
the same Sersic profiles and is not a fit to the data.
{\it Bottom:}
Ratio of FUV emission within one disk scale length \rd\ to that within the
annulus bounded by 1-3\rd\
plotted against \paramn. The higher the ratio $FUV_{1R_D}/FUV_{3-1R_D}$, the
more centrally concentrated
the FUV emission, and hence star formation.
Here we see a correlation in the sense that the more centrally concentrated the
star formation, the higher is \paramn.
The solid line (correlation coefficient 0.61 and standard deviation of 0.4) is 
$n_{HI} = (0.37\pm0.07) + (0.36\pm0.08)\log FUV_{1R_D}/FUV_{3-1R_D}$.
\label{fig-concen}}
\end{figure}

An obvious connection between stars and gas would be that the more centrally
concentrated the gas is, the more centrally concentrated we would expect the stars
and star formation to be. With that in mind, we plot measures of the concentration
of stars and star formation against \paramn\ in Figure \ref{fig-concen}. $C_{31}$, a
measure of the central concentration of the stars, is taken from \citet{zhang12}. It
is defined as the ratio of the radii that encompass 75\% and 25\% of the total
stellar mass  \citep{devauc77}. The larger $C_{31}$ is, the more centrally concentrated
the stellar mass.  Here we see a slight correlation between the central
concentration of stars and the central concentration of gas as described by \paramn.
The dashed line shows what to expect if the surface densities of stars and gas
have the same Sersic profiles, i.e., evaluating $C_{31}$ and $n$ for a single Sersic
function. The approximate agreement with the observations suggests that the central
concentration of stellar mass is related to the central concentration of \HI\ gas.

\begin{figure}[t!]
\epsscale{0.6}
\vskip -0.2truein
\plotone{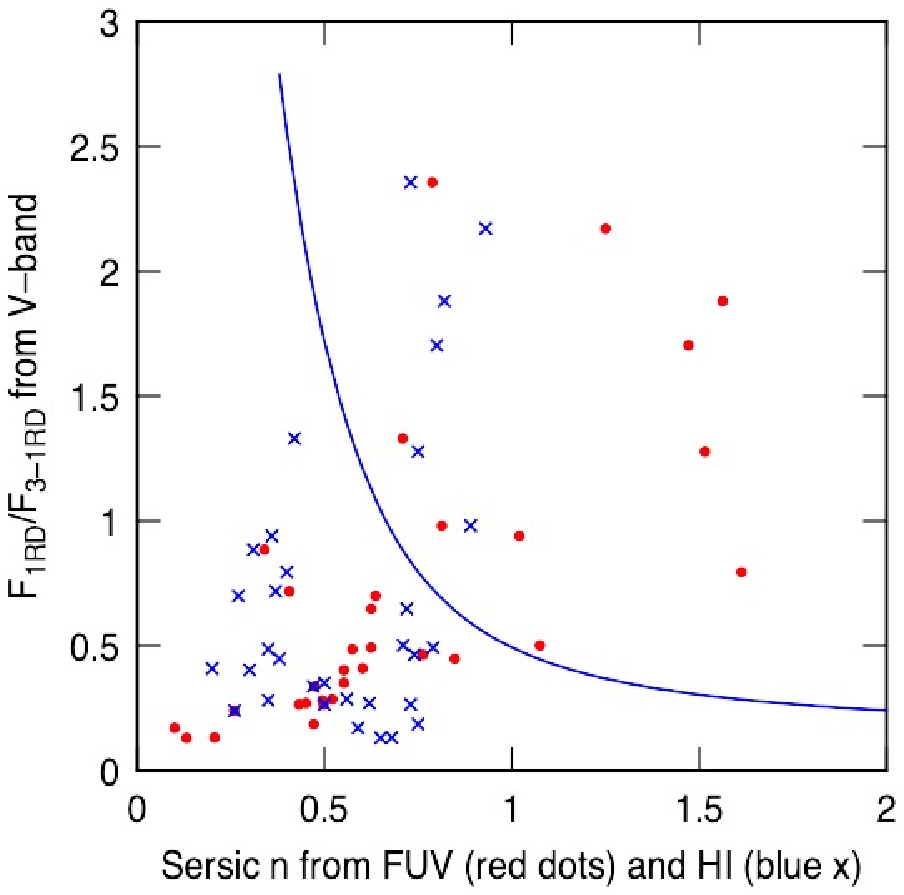}
\vskip -0.3truein
\caption{An examination of the origin of the correlation in the bottom panel
of Fig \ref{fig-concen}. The curve shows the
ratio of the flux inside one scale length to that between 1 and 3 scale lengths
versus the Sersic index for a pure Sersic profile. This curve does not agree with
the observations in the previous figure which mix together all three profiles: FUV, $V$, \HI.
The points use the fitted Sersic profiles for FUV
for each galaxy, with the ratio of the integral of the FUV
flux inside the $V$-band scale length to that between 1 and 3 scale lengths
plotted versus the \HI\ Sersic index (blue x marks)
and the FUV Sersic index (red dots). 
\label{sersic_f1_over_f31}}
\end{figure}

To examine the concentration of star formation, we use the ratio of the FUV
flux in the central scale length to the FUV flux in an annulus around the center
between 1 and 3 \rd.  We denote this ratio by $FUV_{1R_D}/FUV_{3-1R_D}$ in
the bottom panel of Figure \ref{fig-concen}.  The higher the ratio, the more
centrally concentrated is the star formation activity.  The figure indicates that
higher SFR concentration corresponds to higher \paramn. 

However, this bottom panel
contains a mixture of galaxy properties;
it determines the FUV
amounts inside various $V$-band scale lengths as a function of the shape of the \HI\ profile.
We examine this issue further in Figure \ref{sersic_f1_over_f31}.
The decreasing curve in this figure shows the
ratio of the flux inside one scale length to that between 1 and 3 scale lengths
versus the Sersic index for a pure Sersic profile. 
The points in Figure \ref{sersic_f1_over_f31}
are a result of integrating Sersic fits to the
FUV for each galaxy to the $V$-band radius \rd\ and between 1 and 3 \rd\
and plotting the ratio of these versus both the \HI\ Sersic index \paramn\ (blue x
marks) and the FUV Sersic index, \paramnfuv\ (red dots). 
The blue x marks are similar to the points plotted in Figure \ref{fig-concen} determined observationally,
and the distribution of
points from the integrals agrees with the distribution of values from the
observations in Figure \ref{fig-concen}. 
However, the red dots are more self-consistent, because they are the integral of the FUV versus the FUV fit parameter \paramnfuv.
The blue crosses, which are versus \paramn, agree pretty well with the pure FUV red dots 
because the \HI\ \paramn\ and the FUV \paramnfuv\ scale with each other, as shown in Figure \ref{sersic_fuv_vs_hi}.
From this we conclude, that the correlation we see in the bottom panel of Figure \ref{fig-concen} is reasonable even though 
the quantities mix the FUV, $V$, and \HI\ galaxy properties.

\subsection{Flat \HI\ surface density profiles and ``dark'' gas}
\label{sect:dark}

The family of Sersic profiles that fit the \HI\ surface density in LITTLE THINGS
\dirr\ galaxies are generally flatter than exponential because $n_{HI} = 0.2$ to 1 for
the \HI\ and a pure exponential would have $n_{HI}=1$ (see Figure
\ref{fig-sersic}). The \HI\ profiles are also flatter than the FUV surface
brightness profiles (see Figure \ref{sersic_fuv_vs_hi}). Could the flatness of the
\HI\ profile be an indication of the presence of ``dark'' gas, gas that is
molecular, and hence not detected in \HI, but not detected in CO observations yet
either? We suggested there was a significant fraction of H$_2$,
23\% of the gas on average, in our sample of dIrr galaxies on the basis of strong
FUV emission and star formation activity away from the regions where there are
prominent \HI\ clouds \citep{hunter19a}.

\begin{figure}[t!]
\vskip -0.2truein
\plotone{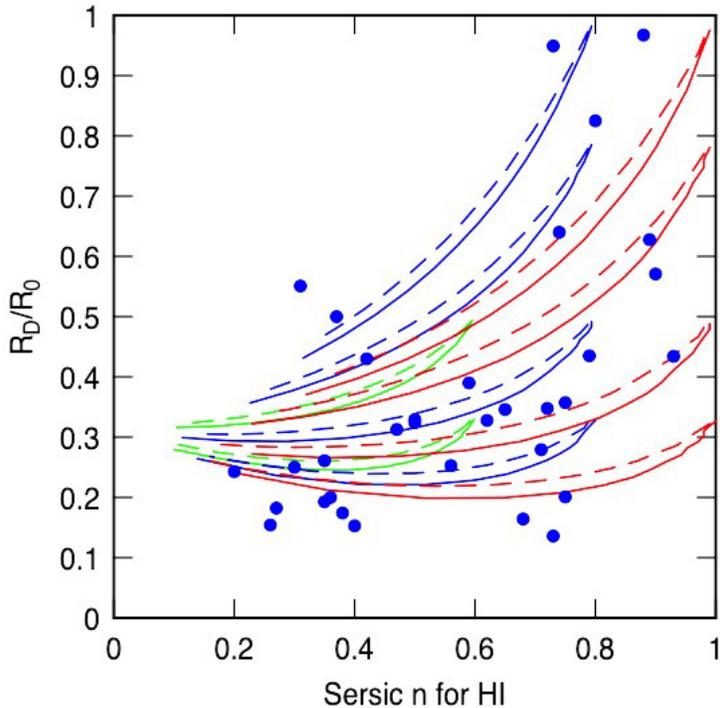}
\vskip -0.25truein
\caption{Comparison of observations of \rd/\paramr\ from the upper right panel of Figure \ref{fig-rat},
plotted as points, with models for total gas Sersic disks that become molecular in the inner
regions. 
The difference between the total and the molecular gas represents \HI,
and this \HI\ is fitted to a Sersic profile with the parameters plotted here.
The molecular fraction of total gas
varies from 5\% at the upper right end of each curve to 100\% at the lower left. The
molecular surface density is assumed to be proportional to the total gas surface density to
the power 1.5 (solid curves) or 2 (dashed curves), in accord with
dynamical models of molecular cloud formation.  
The different colored curves show different intrinsic \paramn\ for the total gas (right-hand limit; red 1, blue 0.8, and green 0.6) and
different intrinsic ratios of \rd\ to \paramr\ (upper limits), increasing
as one goes up the figure.
The models span the range of parameters
given by the observations. This suggests that the Sersic index \paramn\ for \HI\ is lower than
for the optical disk and the scale length for \HI\ is higher than for the optical
disk because of systematic conversions of \HI\ to molecules in the inner regions of these galaxies.
\label{sersic_rstar_over_rhi}}
\end{figure}

We examine this possibility again here in two ways. First, we return to the
top-right panel of Figure \ref{fig-rat} which shows the ratio of the $V$-band scale
length \rd\ to the \HI\ scale length \paramr\ versus the \HI\ Sersic index \paramn. These data points are
reproduced in Figure \ref{sersic_rstar_over_rhi} where we superimpose model
curves that attempt to fit the range of points. The models have molecular fractions
at the galaxy center that range from 5\% to 100\%. For the 5\% fractions, the model
\HI\ profile is essentially the total gas profile, and these models correspond to
the upper right-hand positions of each curve.  The curves then trace down and to the
left as the central molecular fraction increases. Thus when the Sersic index for
\HI\ is 1 on the abscissa, like the $V$-band profile which has an index close to 1,
the two profiles, total gas and $V$-band, have about the same shape. The different
curves that reach $n_{HI}=1$, which are the red curves, show different intrinsic ratios
of $V$-band to \HI\ scale lengths, ranging from 0.33 to 1 as one goes up the figure.
Similarly, the other curves have right-hand limits at the intrinsic \HI\ Sersic
index in the model and they have upper limits at the intrinsic ratio \rd/\paramr.

To trace out the rest of the curves we assume that the H$_2$ surface density is
proportional to the total gas surface density to the 1.5 power (solid curves) or 2.0
power (dashed curves), from the Kennicutt-Schmidt relation that follows from a
dynamical model for molecular cloud formation \citep{elmegreen15,elmegreen18}. In
this model, the rate of conversion of total gas into dense gas, traced by CO or
H$_2$, is the dynamical rate at the midplane density of total gas. For the 1.5
power, the disk thickness varies more slowly with radius than the surface density,
as in several ULIRGS studied by \citet{wilson19} and the main parts of spiral galaxies 
\citep{ee20}.  For the 2.0 power, the gas
disk flares with radius because it is self-gravitating with a nearly constant
velocity dispersion \citep{elmegreen18}.  The \HI\ surface density is then taken to
be the total gas minus the H$_2$.  Sersic fits to this \HI\ surface density are given
by the curves. As the H$_2$ fraction increases, the \HI\ profile flattens, lowering
\paramn, and the \HI\ scale length increases, lowering the ratio $R_{\rm D}/R_{\rm 0,HI}$.
The observations are traced out well by this model.

A second way to test for the presence of molecules is to use the Sersic fit to the
FUV intensity for each galaxy and convert it to a radial profile of SFR density,
$\Sigma_{\rm SFR}$, and then convert this SFR density profile to a molecular density
profile by multiplying it by a constant molecular gas consumption time of
$2\times10^9$ years \citep{leroy08}. According to the dynamical model of star
formation \citep{elmegreen18}, the molecular consumption time is different from the
dynamical time at the midplane, used above for the relation between $\Sigma_{\rm
SFR}$ and total gas, because the consumption time is related to the dynamical time
at the characteristic density of the molecular material (multiplied by the inverse
of some efficiency), which is much higher than the average midplane density except
in starburst galaxies. Whether the molecular material is observed in CO, as for
spirals, or not observed in CO, as for dIrrs, presumably depends on the metallicity,
which is much lower in dIrrs \citep{rubio15}.

The projected SFR density is derived from the projected FUV intensity using the relation:
\begin{equation}
\Sigma_{\rm SFR}=10^{-0.4\mu_{\rm FUV}+7.155}\;M_\odot\;{\rm pc}^{-2}\;{\rm Myr}^{-1},
\end{equation}
which assumes negligible dust extinction, a Chabrier stellar initial mass function \citep{chabrierimf},
and the calibration in \cite{kennicutt98}, modified for sub-solar metallicities by
\cite{hunter10}. The FUV surface brightness in magnitudes per square arcsec is
denoted by $\mu_{\rm FUV}$ and comes from the Sersic fit to the FUV.

\begin{figure}[t!]
\epsscale{1.}
\plotone{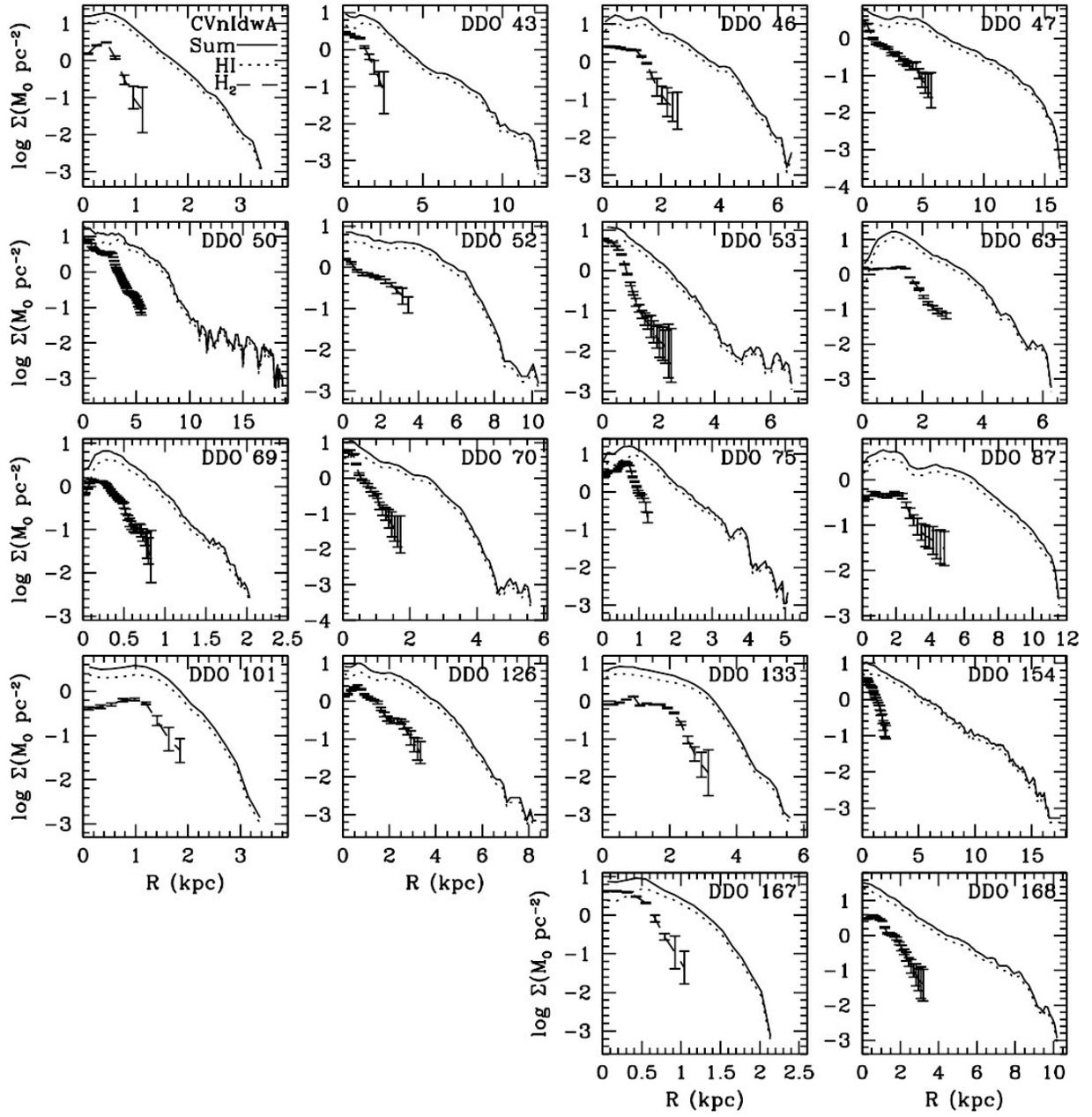}
\caption{Logarithm of the sum \HI$+$He$+$H$_2$ surface density profiles (solid lines), the \HI\ surface density (dotted lines), 
and the H$_2$ surface density profiles inferred from the FUV (dashed lines).
Helium is taken to be 34\% of \HI, and the H$_2$ is inferred from the FUV emission as discussed in the text.
The uncertainties in the H$_2$ profiles are the uncertainties of the logarithm of the FUV flux.
The blanks in the plots are the galaxies without FUV data (DDO 155, DDO 165, IC 10, UGC 8508)
and are maintained to facilitate comparison with Figure \ref{fig-sersicprof}.
\label{fig-totprof}}
\end{figure}

\begin{figure}[h!]
\epsscale{0.5}
\vskip -0.2truein
\plotone{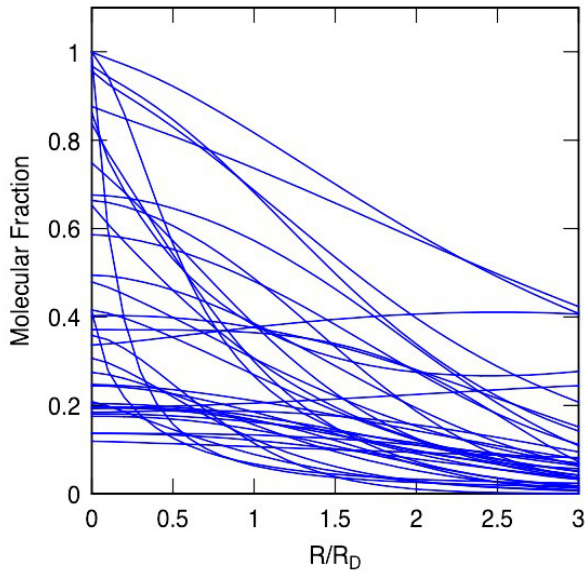}
\vskip -0.2truein
\caption{Radial profiles of model molecular fractions obtained by converting the
FUV surface brightness to molecular surface density and dividing by the sum of this
surface density and the \HI\ surface density.
\label{sersic_molefract}}
\end{figure}

The projected molecular density profile obtained from the FUV intensity in this way is then
added to the observed projected \HI\ Sersic profile to get the total gas total projected gas profile.
All three gas profiles are shown in Figure \ref{fig-totprof}. Figure
\ref{sersic_molefract} shows the radial profiles of the molecular fraction, 
obtained from the ratio of the projected molecular surface density to the total. The
molecular fractions are typically high in the center where the star formation rate
is higher than expected if all of the gas is from the observed \HI\ surface
density. The average molecular fraction for all galaxies, measured out to $3R_{\rm
D}$, is $0.23\pm0.17$, as obtained from the ratio of $\int \Sigma_{\rm H2}2\pi R
dR$ to $\int \Sigma_{\rm sum}2\pi R dR$.

\clearpage

\hspace{-0.85truein} \vspace{-2.truein}
\includegraphics[scale=.9]{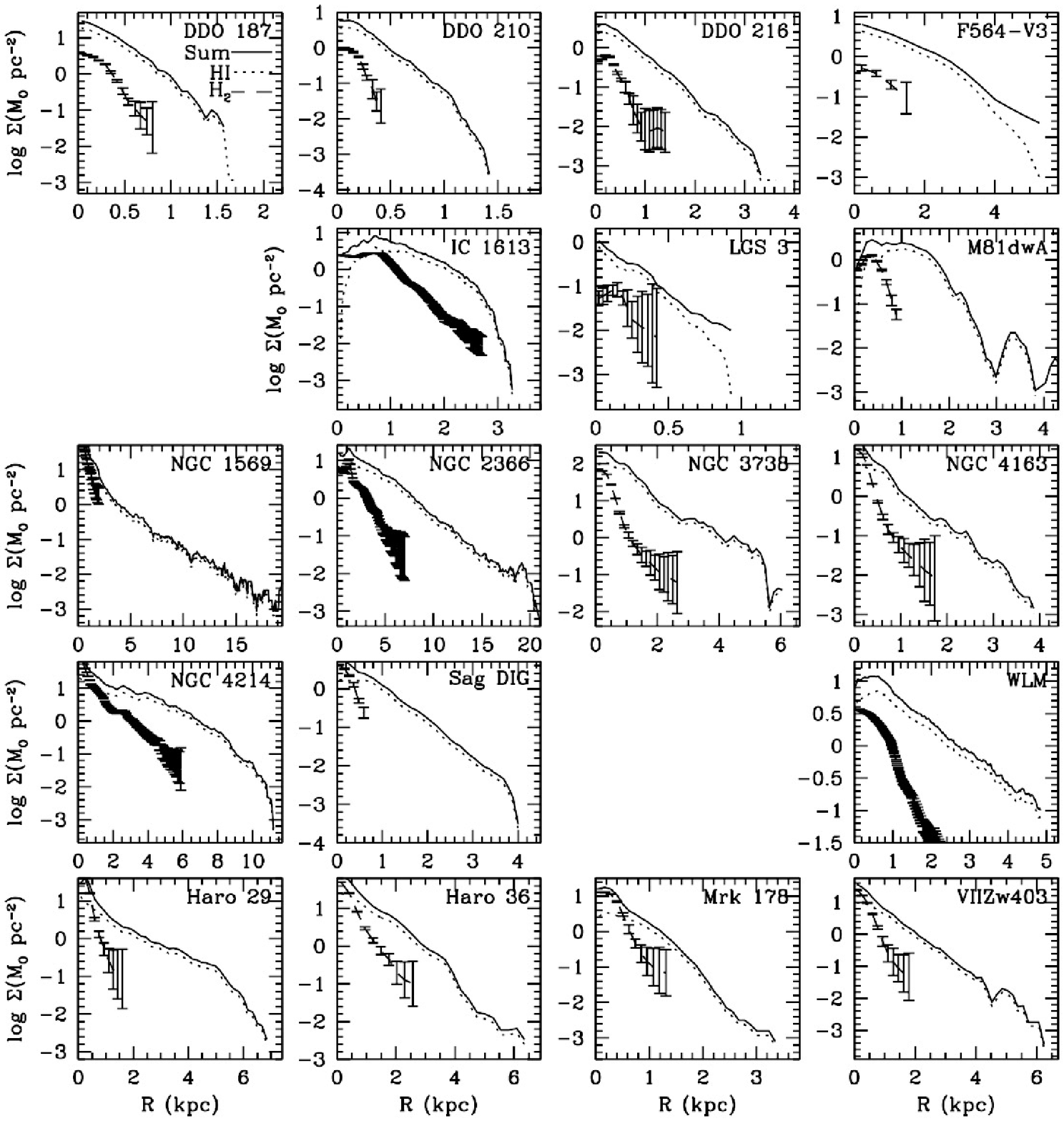}

\clearpage

\section{Investigating \rbr}

A ubiquitous but perplexing feature of stellar radial profiles is a sharp change in the slope of the
exponential fall-off. This is a feature of most exponential disks, both spiral and dwarf irregular.
Here we explore connections between \rbr\ and other radial attributes of the \dirr\ galaxies.

\subsection{Comparison with \HI\ profiles}

\begin{figure}[t!]
\epsscale{0.6}
\plotone{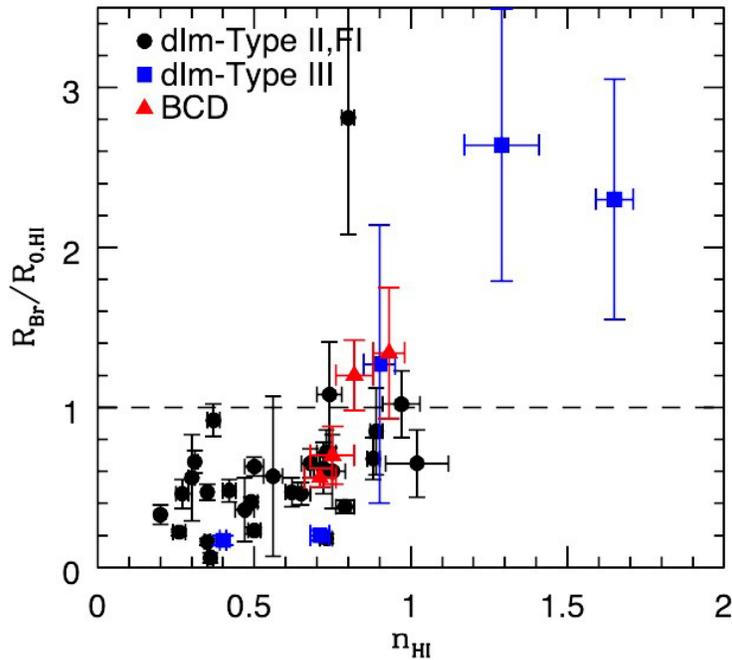}
\caption{
Break radius \rbr\ from the $V$-band surface brightness profile
normalized by \paramr\ plotted against \paramn.
The correlation coefficient is 0.7 and the standard deviation is 0.5.
The dashed horizontal line marks the ratio corresponding to \rbr$=$\paramr.
\label{fig-nrbr}}
\end{figure}

In Figure \ref{fig-nrbr} we plot the radius at which the $V$-band surface brightness profile changes slope \rbr\
against the Sersic parameter \paramn.
The \rbr\ are normalized by $R_0$ from the Sersic fit to the \HI\ surface density profile.
For most galaxies, \rbr\ is smaller than \paramr.
However, there is a modest increase in \rbr\ relative to \paramr\ around $n_{HI}=1$.
The three galaxies with high values of \rbr/\paramr\ also have high uncertainties in this quantity.
The origin of the relationship in Figure \ref{fig-nrbr} is probably similar to
that in the top-left of Figure \ref{fig-rat}, namely, $R_{\rm Br}$ increases with
the size of the galaxy. (See also Section \ref{sec-stellardisk} below
where we show that the ratio of \rbr\ to \rd\ is approximately a constant: \rbr/\rd\ $\sim$ 0.5-2).

\subsection{Comparisons with \HI\ rotation curves}

\begin{figure}[t!]
\epsscale{0.4}
\vskip -0.2truein
\plotone{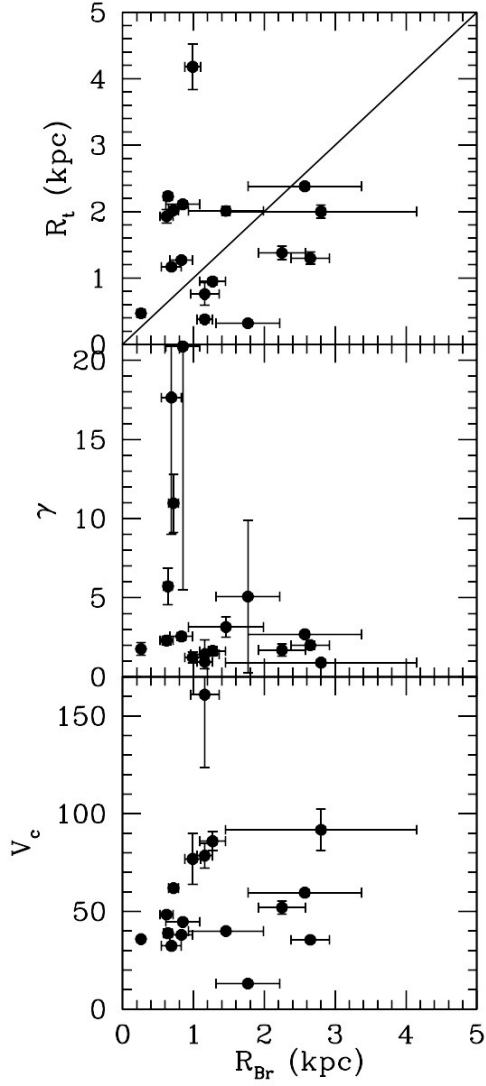}
\vskip -0.2truein
\caption{
Parameters from the fit to the \HI\ rotation curve plotted against the break radius \rbr.
The solid line in the top panel denotes $R_t=R_{Br}$ and is not a fit to the data. 
The correlation coefficient of the data in the top panel is 0.04 and the standard deviation is 0.95.
\label{fig-rotrbr}}
\end{figure}

In Figure \ref{fig-rotrbr} we plot parameters that characterize the fit to the \HI\
rotation curve against the break radius \rbr\ in kiloparsecs. A strong correlation
between \rbr\ and the transition radius $R_t$ or the sharpness of the transition
$\gamma$ would point to an underlying kinematic explanation of the breaks in dwarf
galaxy surface brightness profiles. The solid line in the top panel denotes
$R_t=R_{Br}$. There we see a lot of scatter around the line of equality. The other
two panels do not show correlations. 

A different correlation in Figure \ref{fig-rotrbr} indicates that larger
galaxies, with larger rotation speeds and larger $R_{\rm Br}$, have smoother-rising
rotation curves. This is also evident directly from Figure \ref{fig-rotcurv}.
Generally we consider larger galaxies to be earlier Hubble types with bulges and
more rapidly rising inner rotation curves, which would give them lower $\gamma$.
This is not the case in the dIrr class.  Possibly we are seeing that strong feedback
from early central star formation scatters the central mass and broadens its
concentration more for more massive dIrrs \citep{governato12,elbadry16}, as that
would make the inner rotation curve rise more slowly.

\subsection{Star formation activity interior and exterior to \rbr}

\begin{figure}[t!]
\epsscale{1.0}
\plottwo{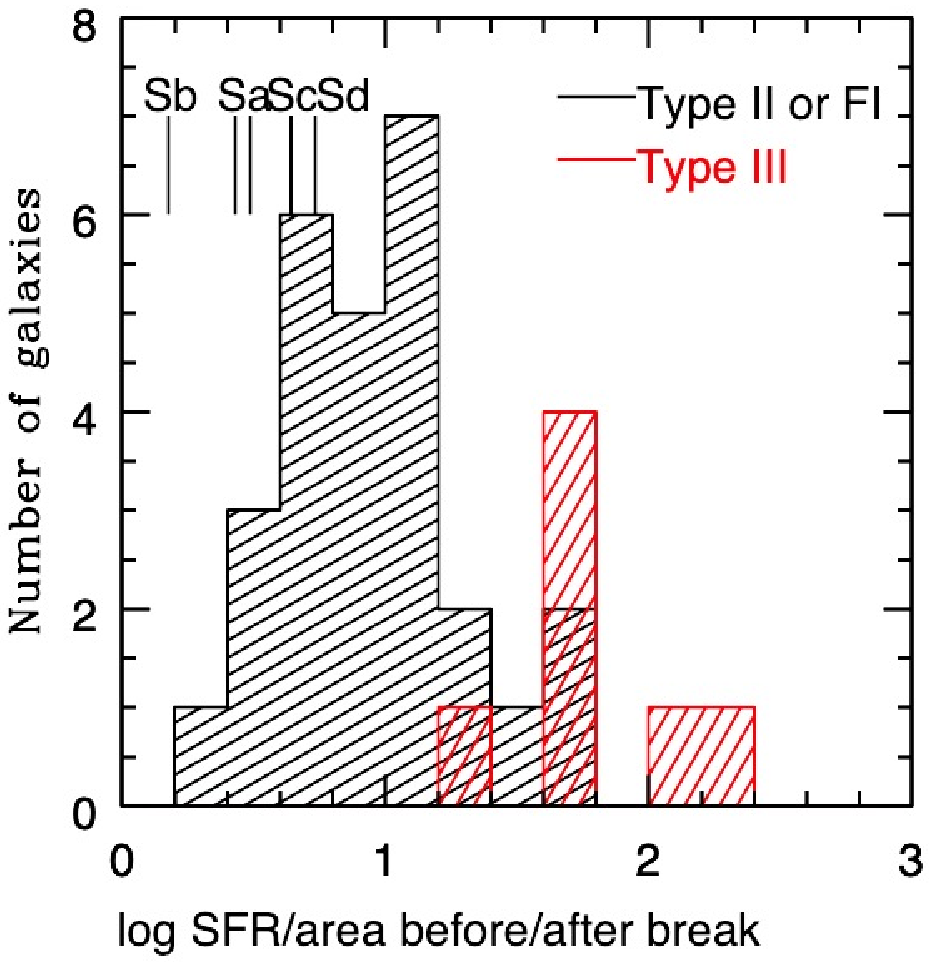}{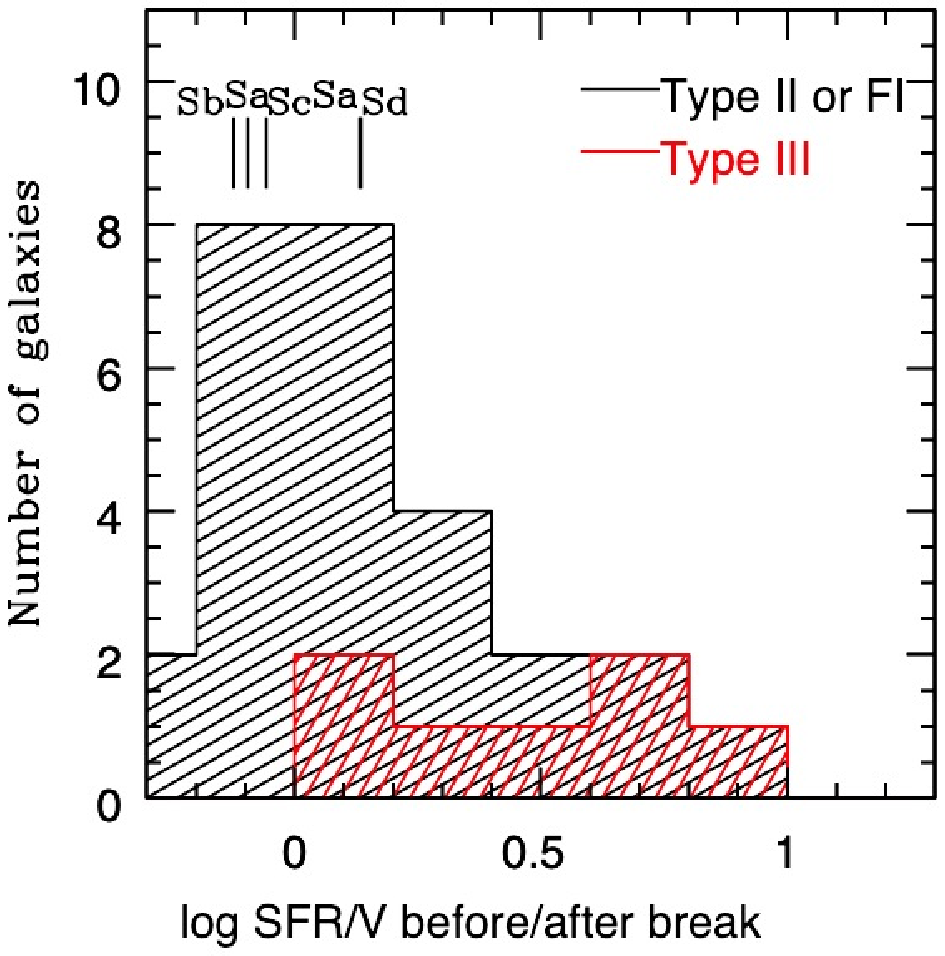}
\caption{
Number of galaxies with given ratios of FUV flux integrated interior to \rbr\ to that exterior to \rbr. Two normalizations are used:
area included in the integration (left) and $V$-band flux in the same region integrated (right).
Values for our spiral galaxy sample are indicated by morphological type of the galaxy, and
Type II or FI \dirr\ are plotted separately from Type III.
\label{fig-fuvrat}}
\end{figure}

In Figure \ref{fig-fuvrat} we show histograms of the ratio of FUV emission interior to \rbr\ to that exterior to \rbr.
The values for the spiral galaxies that were included in this study are marked by spiral morphological type.
Two normalizations of the FUV flux are shown: to the area over which the flux is integrated and
to the $V$-band flux integrated over the same area as for the FUV.
We separate galaxies whose profiles bend downward (Type II and FI) from those that bend upward (Type III).

For the dIrrs the ratios with the $V$-band normalization vary from 0.53 (DDO 87) to 7.5 (NGC 3738) with a median value of 1.4.
However, down-bending types tend to have lower ratios than up-bending types. 
The number of spirals is small and thinly divided by morphological type, but all five have down-bending profiles
and have ratios that are typical of the down-bending dwarfs.
Thus, the down-bending \dirr\ and spirals have approximately the same amount of normalized star formation
interior to \rbr\ as exterior, while the up-bending \dirr\ have more centrally concentrated star formation.
For the area normalization, up-bending types also have higher ratios than down-bending types, but the spirals
tend towards lower values than typical of the dIrrs.
Thus, we see that the star formation activity, relative to the integrated light of older stars, in many \dirr\ galaxies with a down-bending profile
does not change drastically at the break, while those with up-bending profiles have systematically more
star formation interior to the break.
Interestingly, the spirals are similar to those of the majority of the down-bending \dirr\ galaxies, suggesting
that in the down-bending Type II galaxies the star formation process does not change drastically
at the break in either spirals or dwarfs, although a larger sample of spirals would be necessary to 
make this statement stronger.

\begin{figure}[t!]
\epsscale{0.55}
\vskip -0.3truein
\plotone{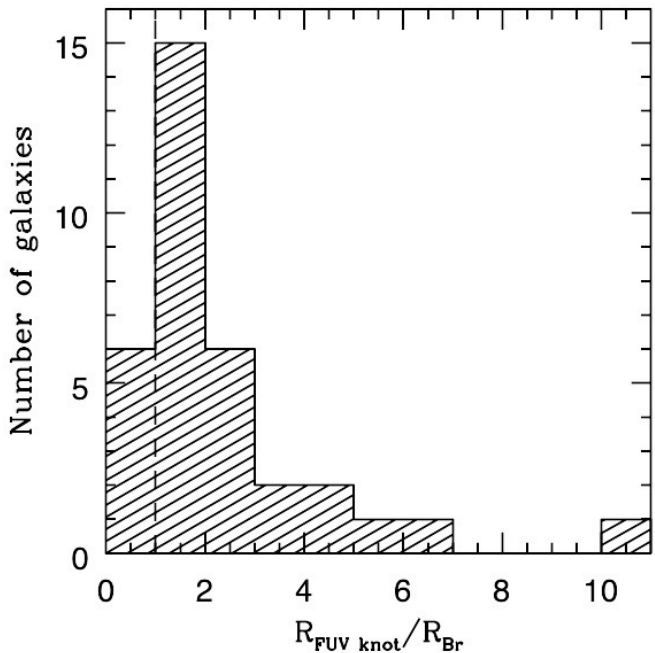}
\vskip -0.3truein
\caption{
Number of galaxies with the given ratio of the radius of the furthest out FUV knot \rfuvknot\ to \rbr.
The vertical dashed line indicates \rfuvknot=\rbr.
\label{fig-outerfuv}}
\end{figure}

In Figure \ref{fig-outerfuv} we show a histogram of the ratio of the radius of the furthest out FUV knot \rfuvknot\ to \rbr.
The median value of the sample is 1.8, so the furthest out FUV knot is beyond \rbr\ in most galaxies.

\subsection{Gas surface density}

\begin{figure}[t!]
\epsscale{0.55}
\vskip -0.3truein
\plotone{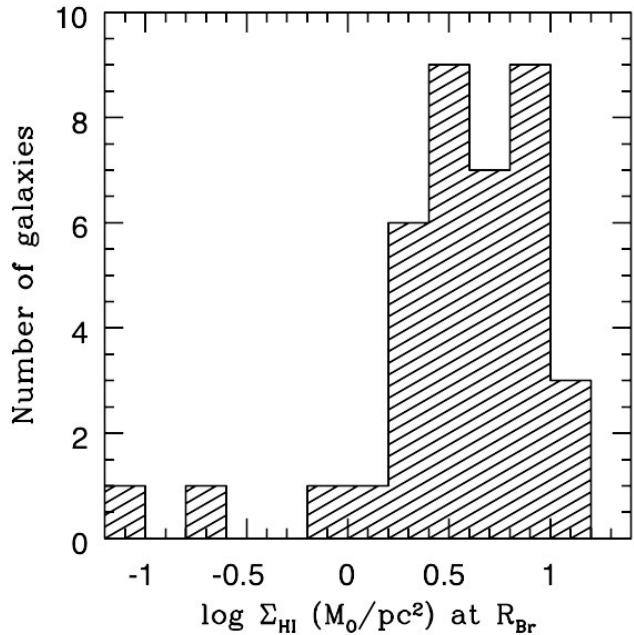}
\vskip -0.3truein
\caption{
Number of \dirr\ galaxies with the indicated \HI\ surface density at \rbr.
\label{fig-hirbr}}
\end{figure}

\citet{schaye04} argues that star formation occurs where there is cold gas that is susceptible to gravitational instabilities, although the connection between cold \HI\ and star formation has been observationally complex 
in \dirr\ \citep[e.g.][]{young96,young03,begum06,deblok06}.
The threshold gas column density for this transition has been argued to be about 3-10$\times10^{20} {\rm cm}^{-2}$
(2.4-8 M\solar pc$^{-2}$).
\citet{outerfuv} found that the furthest out FUV knot in the LITTLE THINGS dIrrs were found above a column density of
2 M\solar pc$^{-2}$.
The break, according to Schaye, occurs where the average gas density drops below this threshold.
In Figure \ref{fig-hirbr} we plot the number of galaxies with a given \HI\ surface density at \rbr. We see that most
galaxies have values between 1.6 and 10 M\solar\ pc$^{-2}$.
Although these values are consistent with various suggestions on density thresholds, this is a very broad range of values,
implying that a single gas density threshold is too simple to explain \rbr.

\subsection{Stellar disk}
\label{sec-stellardisk}

\begin{figure}[t!]
\epsscale{0.6}
\vskip -0.4truein
\plotone{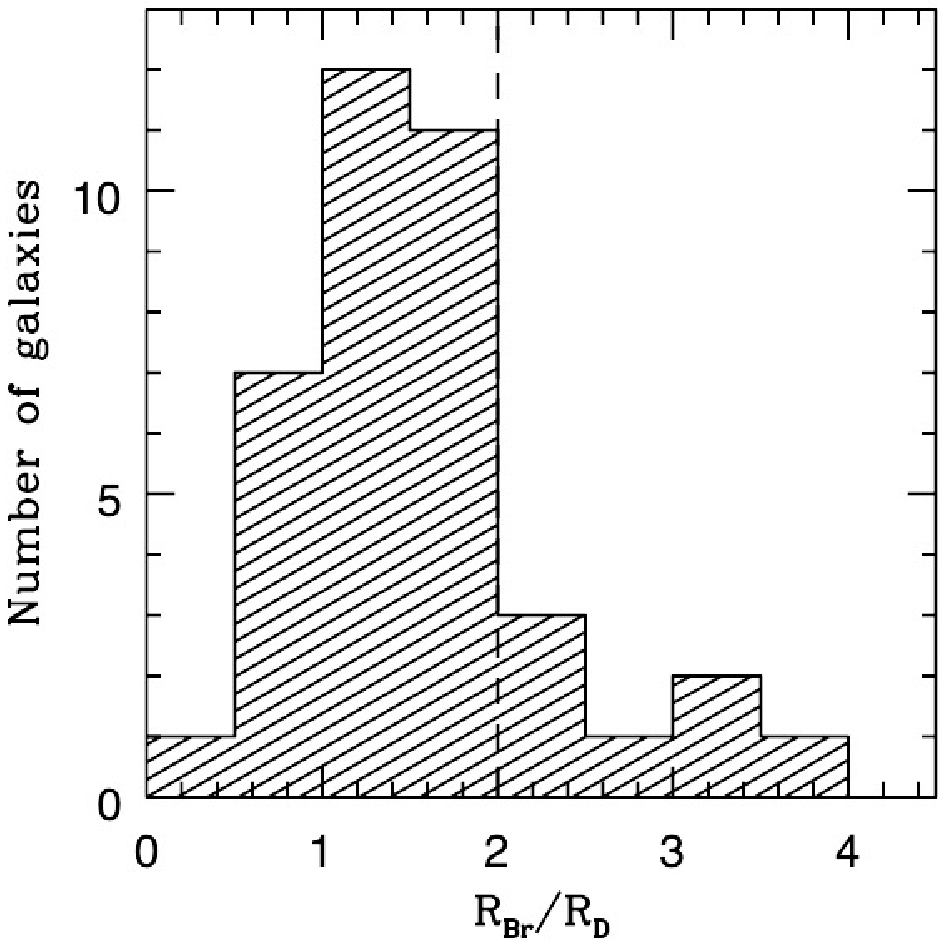}
\vskip -0.3truein
\caption{
Number of \dirr\ galaxies with the indicated \rbr\ in units of \rd.
The vertical dashed line marks \rbr=2\rd.
\label{fig-rbrrd}}
\end{figure}

\citet{li05} found from hydrodynamic simulations that there should be a sharp drop in the SFR at 2\rd.
They suggest that stars are more important than gas in destabilizing dwarf disks, although it is not clear if this means the
actions of the stars or the stellar densities play a key role.
In Figure \ref{fig-rbrrd} we plot a histogram of the ratio \rbr/\rd\ for the LITTLE THINGS galaxies.
We see that the break radius in most (82\%) of these galaxies does occur at 0.5-2\rd.
Furthermore, \citet{herrmann16} shows that \rbr\ is found at a stellar mass surface density of 1-2 M\solar\ pc$^{-2}$
for Type II dIrrs, although higher mass surface densities for BCDs.
Thus, there 
appears to be a relationship between \rbr\ and the stellar surface density.

\subsection{Stellar bar potentials}

\begin{figure}[t!]
\epsscale{0.6}
\vskip -0.25truein
\plotone{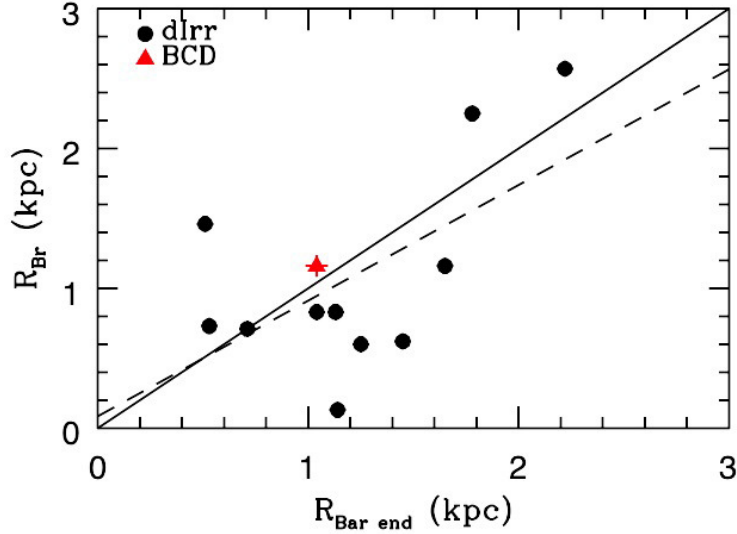}
\vskip -0.25truein
\caption{
Largest distance of the end of the bar from the center of the galaxy \rbar\ vs.\ \rbr\
for the LITTLE THINGS galaxies with bars \citep{he06}.
The slanted solid line denotes a one-to-one relationship.
Errorbars are plotted but are generally smaller than the point size.
The dashed line is a fit to the points, with correlation coefficient of 0.6 and standard deviation of 0.6:
$R_{Br} = (0.086\pm0.46) + (0.83\pm0.35)R_{Bar end}$.
\label{fig-bars}}
\end{figure}

Some \dirr\ galaxies are barred as evidenced by the rotation of the optical surface brightness isophotes with increasing radius.
In one case, we also observe streaming motions of the \HI\ around the bar \citep{hunter19b}.
There are 12 LITTLE THINGS dIrrs for which there is evidence for a bar \citep{he06},
and we use these to explore the impact of a bar potential on \rbr.
For these galaxies we have determined the distance of the end of the bar in the plane of the galaxy.
Since most
bars are offset from the galactic center, we have found the furthest point of the bar from the center of the galaxy.
This information is given in Table \ref{tab-bars}.
$R_{\rm Bar}$ is the semi-major axis of the bar and \rbar\ is the largest distance of the
edge of the bar from the center of the galaxy.
\rbr\ is plotted against \rbar\ in Figure \ref{fig-bars}.
We see that \rbr$\sim$\rbar, which could imply a connection between the bar and the profile break in
these galaxies. However, not all \dirr\ with breaks have bars.

\begin{deluxetable}{lccc}
\tabletypesize{\scriptsize}
\tablecaption{Stellar bar structures \label{tab-bars}}
\tablewidth{0pt}
\tablehead{
\colhead{Galaxy}
& \colhead{$R_{\rm Bar}$ (arcsec)}
& \colhead{\rbar\ (kpc)}
& \colhead{\rbr/\rbar} \\
}
\startdata
DDO 43      &  $13.5\pm0.7$   & $0.51\pm0.03$ &  $2.9\pm0.16$ \\
DDO 70      &  $102.9\pm5.1$ & $1.14\pm0.03$ &  $0.1\pm0.01$ \\
DDO 126    &  $35.9\pm1.8$   & $1.25\pm0.04$  & $0.5\pm0.03$ \\
DDO 133    &  $83.9\pm4.2$   & $1.78\pm0.07$  & $1.3\pm0.07$ \\
DDO 154    &  $54.3\pm2.7$   & $1.45\pm0.05$  &  $0.4\pm0.03$ \\
F564-V3     &  $11.4\pm0.6$   &  $0.53\pm0.02$  & $1.4\pm0.09$ \\
NGC 2366  &  $119.3\pm6.0$ &  $2.22\pm0.10$  & $1.2\pm0.05$ \\
NGC 3738  &  $64.8\pm3.2$   &  $1.65\pm0.08$  & $0.7\pm0.03$ \\
NGC 4163  &  $45.6 \pm2.3$. &  $0.71\pm0.03$  & $1.0\pm0.06$ \\
NGC 4214  &  $53.9\pm2.7$   &  $1.04\pm0.04$  & $0.8\pm0.03$ \\
WLM          &   $203.1\pm10.1$  & $1.13\pm0.05$  & $0.7\pm0.04$ \\
Haro 36      &  $17.7\pm0.9$   & $1.12\pm0.04$   & $1.0\pm0.04$ \\
\enddata
\end{deluxetable}

\clearpage

\section{Summary}

We have examined the relationship between properties of the stellar disk, the gas
disk, and young stars in the LITTLE THINGS sample of nearby \dirr\ galaxies. The
stellar disk is characterized by the disk scale length \rd, the radius at which the
$V$-band surface brightness profile changes slope \rbr, and a measure of the central
concentration of the stellar mass $C_{31}$. The \HI\ surface density radial profile
is fit with a Sersic function with parameters of extrapolated central surface gas density \paramhi,
characteristic radius \paramr, and curvature of the profile \paramn. The FUV surface
brightness profile is fit with three similar parameters: \paramfuv, \paramrfuv, \paramnfuv.

We include the ratio of the radius that contains 50\% of the \HI\ to the radius that
contains 90\% of the total \HI. The rotation curve is fit with a function that
includes the asymptotic velocity
\paramvc, the radius where the rotation speed levels off or increases more slowly
\paramrt, and the sharpness of the transition
\paramgamma. FUV images are used as tracers of young stars, and we specifically look
at the ratio of young to older stars interior and exterior to \rbr. We include 5
spiral galaxies in this examination for comparison. We also consider the radial
extent of \ha\ emission \rha, the radial extent of knots of FUV emission \rfuvknot, the
ratio of FUV emission within one disk scale length to that in an annulus from 1 to 3
\rd\ as a measure of the concentration of the star formation activity, and the
galactic SFR determined from the integrated FUV emission. We also compare our data
to predictions for what happens at \rbr, including the gas and stellar mass surface
densities and \rd.

Comparing the \HI\ disk with the stellar disk, we find the following:

(1) Most of our dIrrs have \HI\ surface density profiles that fall off with radius with smaller \paramn\ than that of an exponential disk. This means that
the \HI\ profile is flatter in the center before falling off more precipitously compared to an $n=1$ exponential disk profile that falls off steadily from the center.

(2) Those galaxies closer to \paramn$=1$ have, relative to \paramr, larger \rha, \rd, and $R_{50}$.
$R_0/R_{50}$ and $R_{50}/R_{90}$ are related to \paramn\ as expected for a Sersic profile.

(3) The integrated SFR increases with extrapolated central \HI\ surface density.

(4) There is no correlation between the \HI\ surface density shape defined by \paramn\ and the degree of central concentration
of the stellar mass, but
the young stars are more centrally concentrated in galaxies with a more
steady fall off of gas density from the center (larger \paramn).

Looking at \rbr, we find the following:

(1) The break radius \rbr\ is generally smaller than the characteristic radius of the \HI\ profile \paramr,
is found at 0.5-2\rd, and
is roughly near the transition radius of the \HI\ rotation curve \paramrt.

(2) For \dirr\ galaxies with down-bending surface brightness profiles,
the ratio of the SFR before \rbr\ to that after the break, normalized to the starlight from older stars in the same area,
is about one. That is, the star formation activity does not change drastically at the break.
A small sample of spirals have similar ratios, suggesting that this applies to spirals as well.

(3) There is a tighter relationship between \rbr\ and the stellar disk than with the \HI\ disk.

Considering the fall-off of \HI\ in the inner regions as indicated by a low
Sersic index, \paramn, we suggest the following interpretation:

(1) The observed increase in \HI\ scale length with decreasing index \paramn\ fits well
to a model 
where the total gas has a Sersic
profile with an index and scale length comparable to that of the stellar disk, and
where there is a range for the molecular fraction in the center that extends up to
100\%. In this model, the molecular surface density scales with a power of the total
gas surface density, consistent with molecular cloud formation at the dynamical rate
of the midplane gas. Conversion of \HI\ to invisible molecules then causes the inner
fall off in the \HI\ profile that increases the \HI\ scale length and lowers \paramn.

(2) The radial profile of the molecular fraction is determined by
converting the radial profile of the FUV flux, fitted with a Sersic
function, into a radial profile of molecular surface density, using standard
calibrations for star formation. The sum of this molecular surface density and the
Sersic fit to the \HI\ profile gives the total gas profile, which then gives the molecular fraction. 
The average molecular fraction in the inner
$3R_{\rm D}$ for all of our galaxies is $23\pm17$\%.


\acknowledgments

A.\ I.\ E., H.\ T., and E.\ G.\  appreciate the MIT Department of Earth, Atmospheric, and Planetary Sciences for supporting the
MIT Field Camp at Lowell Observatory in 2014, 2017, and 2019 and Dr.\ Amanda Bosh for organizing and running that program.
B.\ B.\ is grateful for funding from the NAU Space Grant program in 2015 and Kathleen Stigmon for running that.
D.\ A.\ H.\ appreciates assistance for publication provided by the National Science Foundation grant AST-1907492.
S.-H.\ O.\ acknowledges support from National Research Foundation of Korea (NRF) grant 
NRF-2020R1A2C1008706 funded by the Korean government Ministry of Science and ICT (MSIT).
We appreciate thoughtful and detailed suggestions from an anonymous referee that helped improve the manuscript.
Lowell Observatory sits at the base of mountains sacred to tribes throughout the region. 
We honor their past, present, and future generations, who have lived here for millennia and will forever call this place home.

Facilities: \facility{VLA} \facility{GALEX}

\appendix

\section{Spiral Surface Photometry}

$UBVJHK$ and \ha\ surface photometry for the LITTLE THINGS dIrr galaxies was presented originally by \citet{he04} and \citet{he06}, 
and other passbands were added as they became available, for example FUV and NUV by \citet{hunter10} and {\it Spitzer} 3.6 and 4.5 micron
by \citet{zhang12}. These azimuthally-averaged profiles were analyzed for breaks by \citet{herrmann13}. 
Measuring the surface photometry on the $V$ and FUV images of spirals followed the same process
as for the dwarfs with the following exceptions: 1) there was no nuclear region in the dIrrs that needed to be subtracted from the profile,
and 2) the reddening correction in the dIrrs was a simple constant and not a function of radius. 
In addition, the process of fitting the profiles and determining the number and location of breaks in the profiles in the dwarfs
was more sophisticated than what we did for the spiral galaxies here. There we were fitting 11 passbands in 141 galaxies and
so doing this entirely by hand was not feasible. So \citet{herrmann13}
wrote an iterative program to determine the best fit and whether there was a single or double
exponential, but some human intervention was required so we referred to this as ``human-assisted computer break fitting.''
See  \citet{herrmann13} for the details.
For the spiral galaxies, the break was determined by eye and a linear fitting algorithm was used for each piece of the profile.
The various tables in this paper will be available in machine readable form when the paper is published.

Note: The radius $R$ in surface photometry profiles refers to the semi-major axis at the mid-point of the annulus in which the
surface brightness or surface mass density was measured.

Here we describe the reduction of the spiral galaxy imaging data.
The spiral sample was chosen to be representative of morphological type Sa, Sb, Sc, and Sd, to not be too edge on,
to be observable by us, to have FUV imaging in the NGS catalog, and to have a break in the $V$-band surface
brightness profile. The sample size was limited by our time, but it gives a suggestion of how spirals might compare to \dirr\ galaxies in
terms of what happens to the star formation at the $V$-band break.
We co-added the $V$-band images, removed foreground and background objects, and fit and subtracted the sky.
We fit an outer contour with an ellipse to determine the center of the galaxy, position angle P.A., and minor-to-major axis ratio $b/a$.
These parameters were held fixed as we measured the $V$-band and FUV flux in ellipses of increasing major axis.
From this we determined the surface brightness in annuli.

We applied a correction for foreground reddening $E(B-V)_{\rm f}$ using $A_V = E(B-V)_{\rm f} \times 3.1$
and $A_{FUV} = E(B-V)_{\rm f} \times 8.24$.
We corrected for internal extinction $E(B-V)_{\rm i}$ using the procedure outlined by \citet{spirals}.
We use the \ha\ extinction as a function of radius from \citet{prescott07}, the ratio of the \ha\
extinction to the reddening from \citet{calzetti00}, and
the relationship of the \ha\ extinction to the extinction of stars from \citet{calzetti97}.

We identified the break in the $V$-band surface brightness profile, and fit the profile interior and exterior to \rbr\ with a straight line.
We used the extinction corrected photometry to determine the total FUV and $V$-band flux interior and exterior to \rbr.
We then subtracted the $V$-band flux of the nucleus from the integrated interior $V$ flux by extrapolating the interior exponential disk
inward.

\begin{figure}
\epsscale{1.1}
\plotone{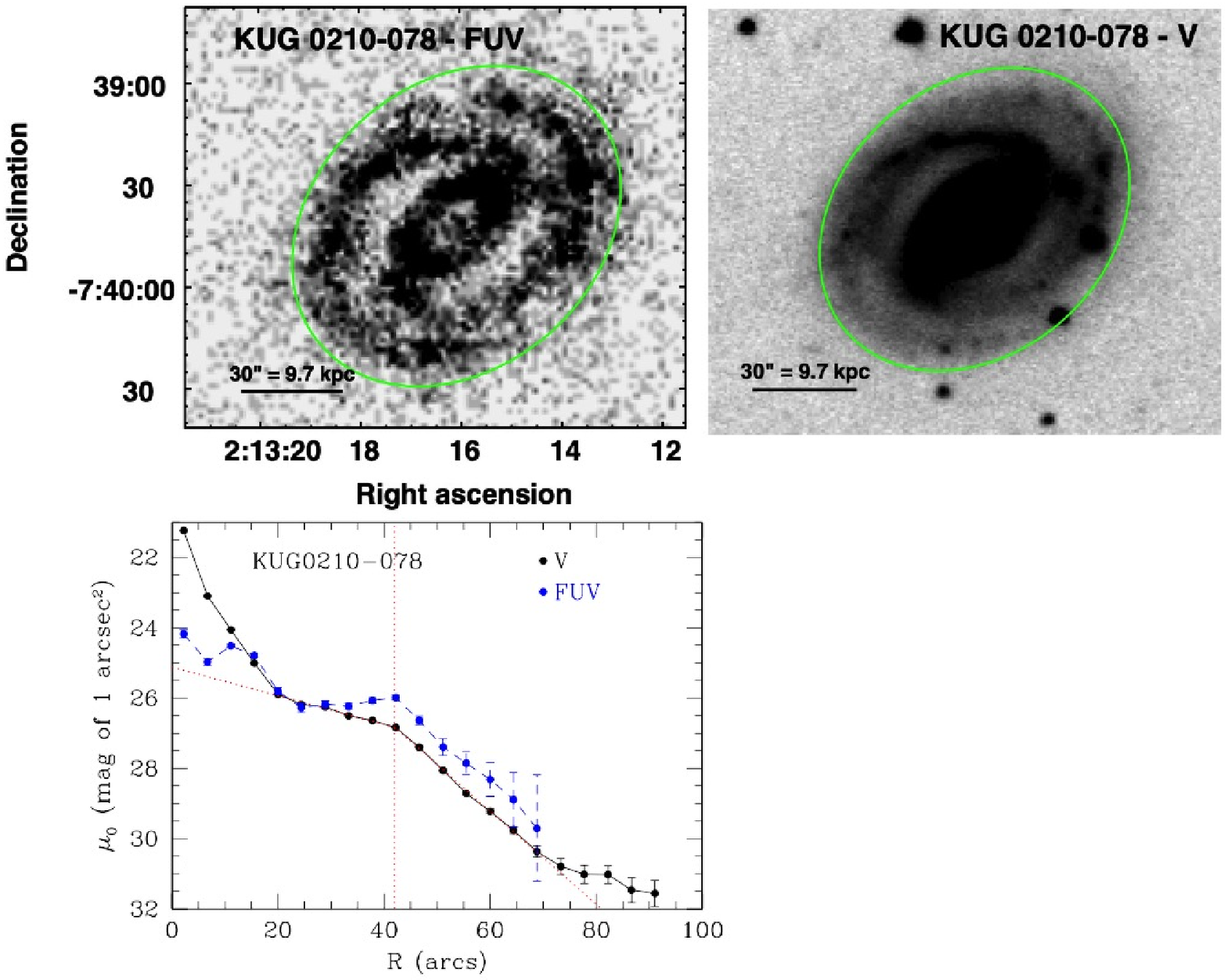}
\caption{
{\it Top:} FUV and $V$-band images of KUG 0210-078. The green ellipse is the 11th ellipse in the surface photometry
and shows the centering and P.A.\ of the ellipses.
{\it Bottom:} FUV and $V$-band surface photometry corrected for foreground and internal extinction.
The vertical red dotted line marks the break radius.
The slanted dotted red lines are the fits to the surface photometry.
\label{fig-kug0210}}
\end{figure}

\begin{figure}
\epsscale{1.1}
\plotone{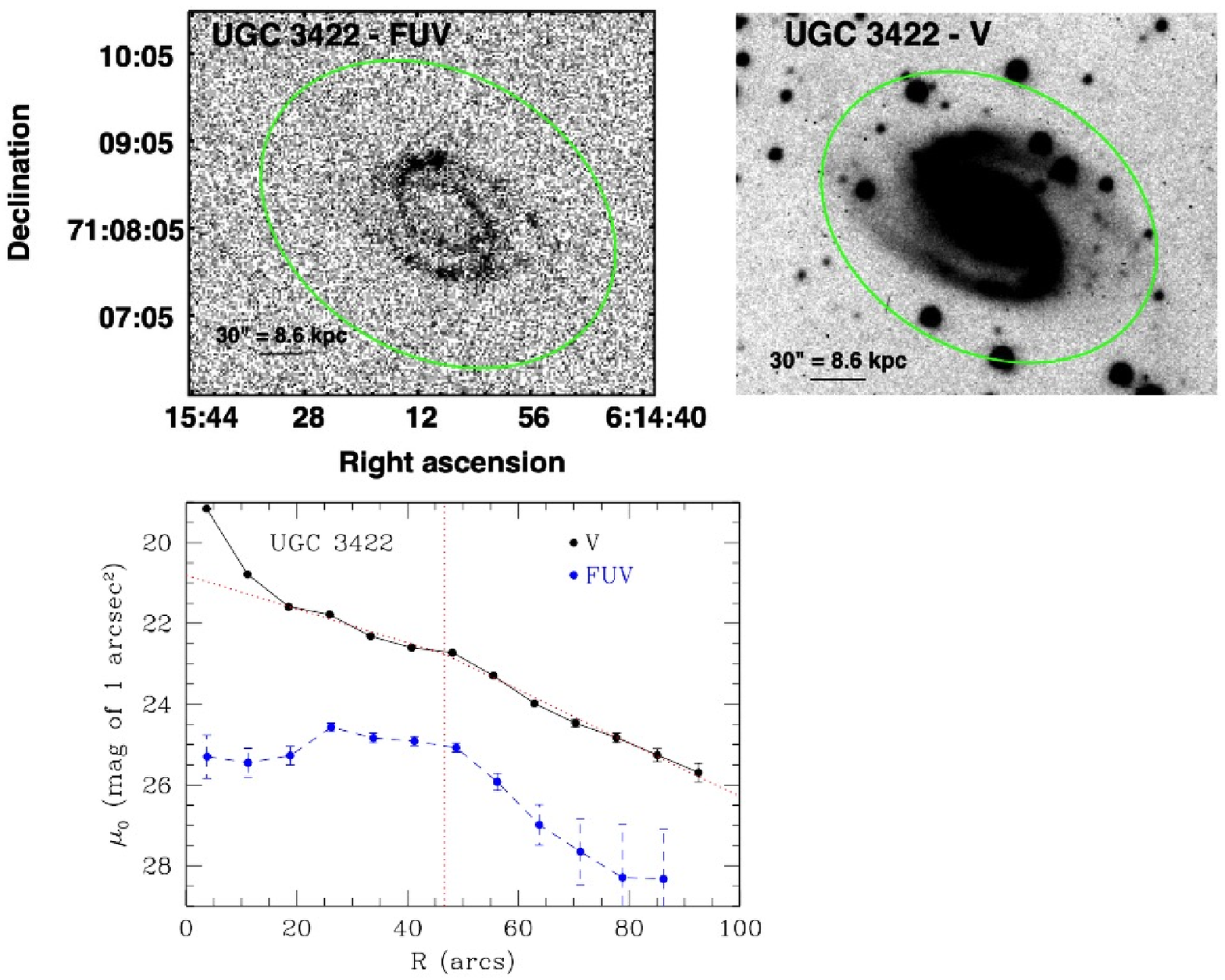}
\caption{
{\it Top:} FUV and $V$-band images of UGC 3422. The green ellipse is the 13th ellipse in the surface photometry
and shows the centering and P.A.\ of the ellipses.
{\it Bottom:} FUV and $V$-band surface photometry corrected for foreground and internal extinction.
The vertical red dotted line marks the break radius.
The slanted dotted red lines are the fits to the surface photometry.
\label{fig-u3422}}
\end{figure}

\begin{figure}
\epsscale{1.1}
\plotone{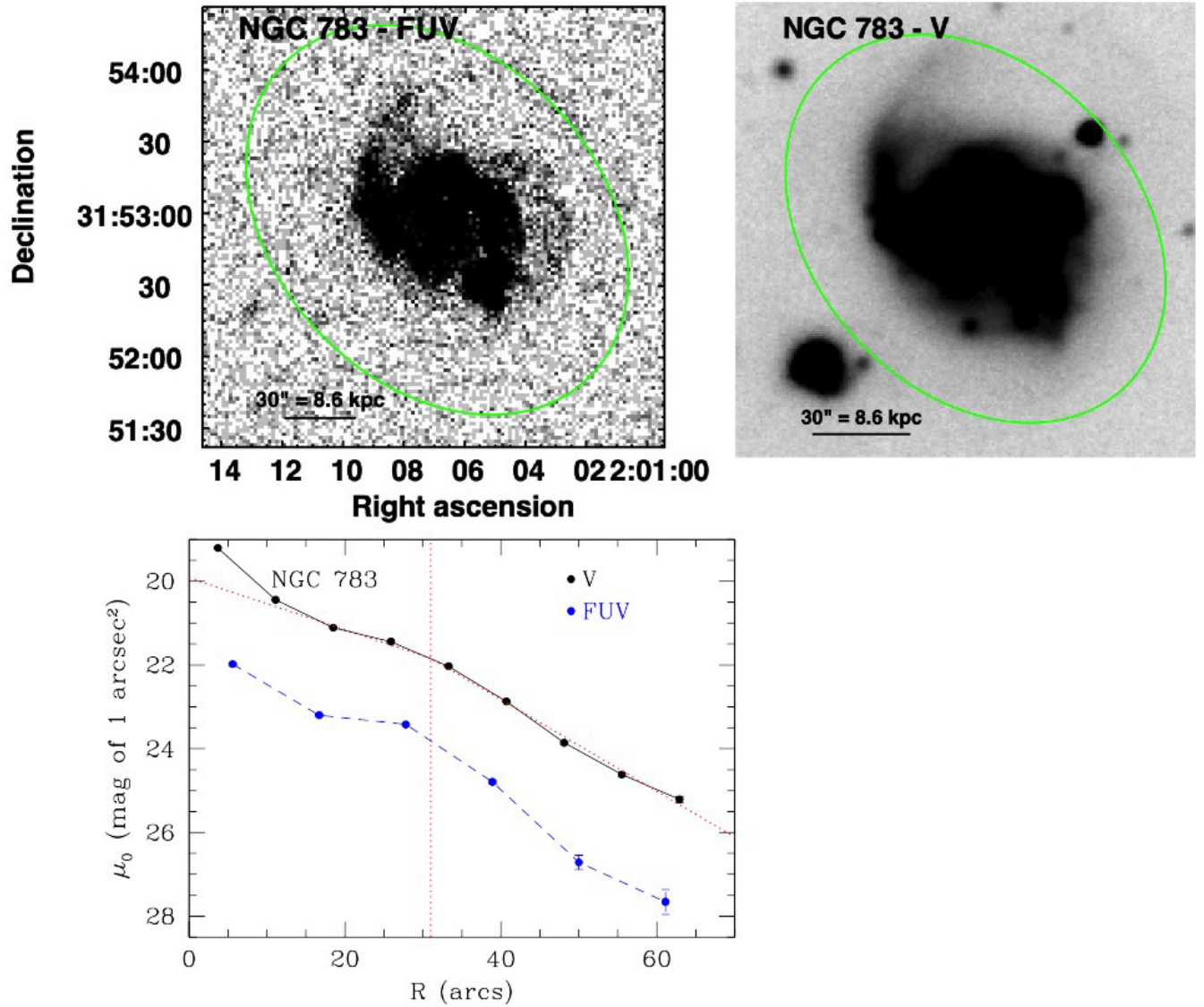}
\caption{
{\it Top:} FUV and $V$-band images of NGC 783. The green ellipse is the 6th ellipse in the surface photometry
of the FUV image and the 9th ellipse in the $V$-band surface photometry,
and shows the centering and P.A.\ of the ellipses.
{\it Bottom:} FUV and $V$-band surface photometry corrected for foreground and internal extinction.
The vertical red dotted line marks the break radius.
The slanted dotted red lines are the fits to the surface photometry.
\label{fig-n783}}
\end{figure}

\begin{figure}
\epsscale{1.1}
\plotone{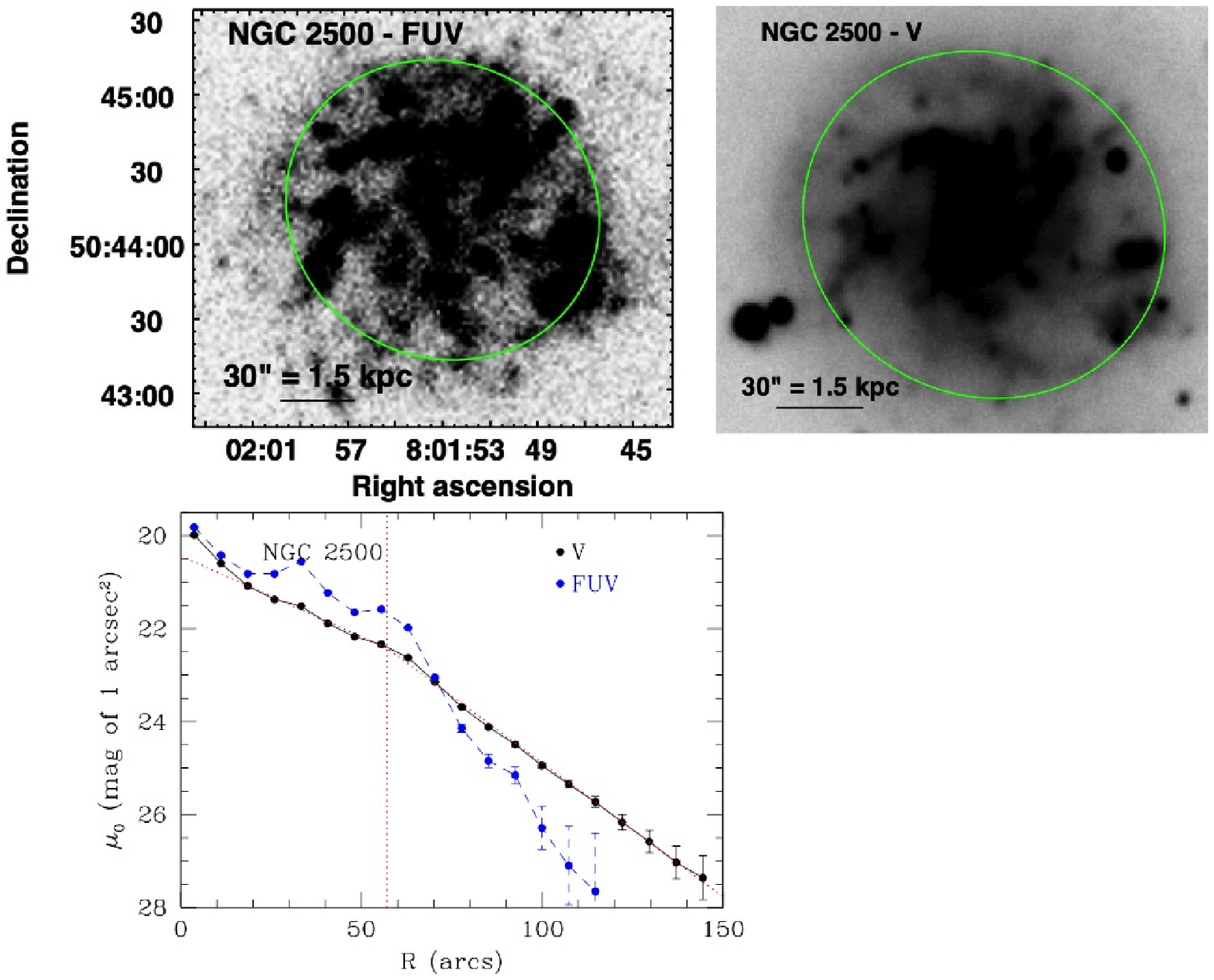}
\caption{
{\it Top:} FUV and $V$-band images of NGC 2500. The green ellipse is the 9th ellipse in the surface photometry
and shows the centering and P.A.\ of the ellipses.
{\it Bottom:} FUV and $V$-band surface photometry corrected for foreground and internal extinction.
The vertical red dotted line marks the break radius.
The slanted dotted red lines are the fits to the surface photometry.
\label{fig-n2500}}
\end{figure}

\begin{figure}
\epsscale{1.1}
\plotone{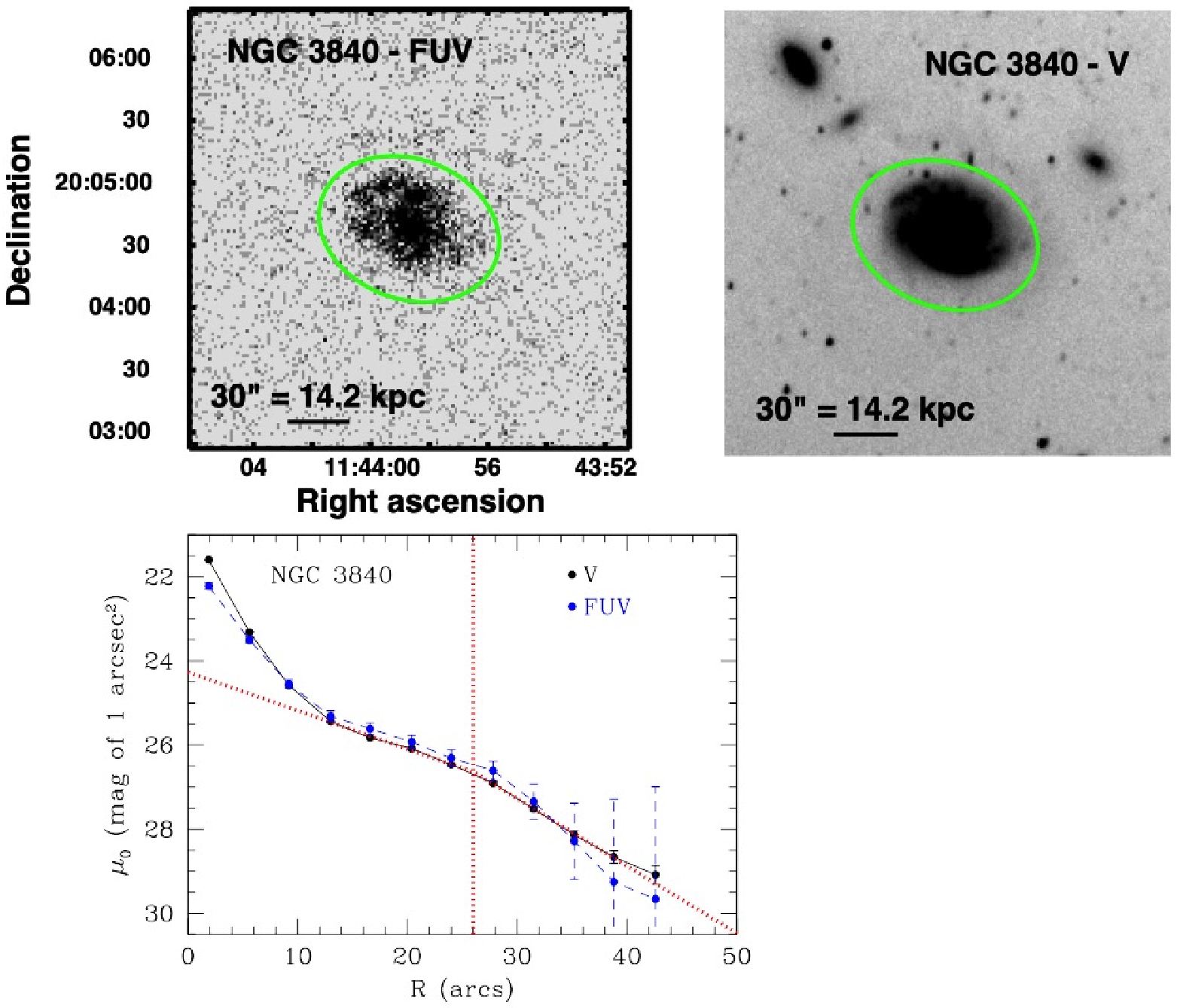}
\caption{
{\it Top:} FUV and $V$-band images of NGC 3840. The green ellipse is the 12th ellipse in the surface photometry
and shows the centering and P.A.\ of the ellipses.
{\it Bottom:} FUV and $V$-band surface photometry corrected for foreground and internal extinction.
The vertical red dotted line marks the break radius.
The slanted dotted red lines are the fits to the surface photometry.
\label{fig-n3840}}
\end{figure}

The $V$ and FUV surface brightness profiles and the exponential fits are shown in Figures \ref{fig-kug0210} to \ref{fig-n3840}.
The galaxies and their properties are given in Tables \ref{tab-spirals}, \ref{tab-spiralprof}, and \ref{tab-ratsspirals}.

\begin{deluxetable}{lccccccc}
\tabletypesize{\scriptsize}
\tablecaption{Spirals included in the FUV study \label{tab-spirals}}
\tablewidth{0pt}
\tablehead{
\colhead{}
& \colhead{}
& \colhead{}
& \colhead{}
& \colhead{}
& \colhead{} \\
\colhead{Galaxy}
& \colhead{Type}
& \colhead{D(Mpc)\tablenotemark{a}}
& \colhead{Ref for D}
& \colhead{M$_V$}
& \colhead{$E(B-V)_{\rm f}$\tablenotemark{b}}
& \colhead{\rbr\ (arcs)\tablenotemark{c}}
& \colhead{\rd\ (arcs)\tablenotemark{d}} \\
}
\startdata
KUG 0210-078 & Sa & 66.9$\pm$4.7   &  NED & $-18.10\pm0.15$ & 0.024   & $41.9\pm2.3$   & 26.5$\pm$1.9 \\
NGC 3840        & Sa & 97.5$\pm$3.1   &   1      &  $-18.45\pm0.07$  & 0.019 & $26.0\pm3.3$ & 14.5$\pm$1.1  \\
UGC 3422        & Sb & 58.8$\pm$9.9   &   2      & $-21.34\pm0.37$ & 0.174   & $46.7\pm16.4$ & 25.9$\pm$3.1 \\
NGC 783          & Sc & 59.4$\pm$5.2   &   3      & $-21.27\pm0.19$ & 0.054   & $31.0\pm3.6$   & 17.5$\pm$2.8 \\
NGC 2500        & Sd & 10.1$\pm$1.4   &   4      & $-18.19\pm0.30 $ & 0.036  &  $57.0\pm5.8$  & 31.0$\pm$0.9 \\
\enddata
\tablenotetext{a}{Distance to the galaxy.
We used a SNIa, SNII, or Tully-Fisher derived distance, when available. Uncertainty in the distance is
folded into the uncertainty of M$_V$.}
\tablenotetext{b}{Foreground Milky Way reddenings $E(B-V)_{\rm f}$ are taken from NED \citep{foregroundred}.}
\tablenotetext{c}{Break radius at which the $V$-band surface brightness profile changes slope.}
\tablenotetext{d}{Disk scale length measured from the $V$-band surface brightness profile.}
\tablerefs{
(1) \citet{springob2009};
(2) \citet{theureau2007};
(3) \citet{ganeshalingam2013};
(4) \citet{tullyandfisher88}}
\end{deluxetable}

\begin{deluxetable}{lccccccccc}
\tabletypesize{\scriptsize}
\tablecaption{Spiral Imaging Photometry \label{tab-spiralprof}}
\rotate
\tablewidth{0pt}
\tablehead{
\colhead{}
& \colhead{$V$ Obs}
& \colhead{}
& \colhead{Center RA, Decl.}
& \colhead{}
& \colhead{}
& \multicolumn{2}{c}{Interior\tablenotemark{a}}
& \multicolumn{2}{c}{Exterior\tablenotemark{a}} \\
\colhead{Galaxy}
& \colhead{(No.\ exposures$\times$exposure time (s))}
& \colhead{FUV exp time (s)}
& \colhead{h:m:s, d:m:s}
& \colhead{P.A. (deg)}
& \colhead{$b/a$}
& \colhead{$\mu_{\rm cen}$}
& \colhead{$b$}
& \colhead{$\mu_{\rm cen}$}
& \colhead{$b$} \\
}
\startdata
KUG 0210-078 & 10$\times$600 & 1680  & 2:13:15.7, $-$7:39:42 & $-48$ & 0.79 & 25.1$\pm$0.1 & 0.041$\pm$0.003 & 21.3$\pm$0.1 & 0.133$\pm$0.002 \\
NGC 3840        & 13$\times$600 &   942  & 11:43:59.0, 20:04:38  &  70      & 0.76 & 24.7$\pm$0.1 &  0.075$\pm$0.006 & 22.6$\pm$0.1 & 0.158$\pm$0.003 \\
UGC 3422        &  9$\times$600  & 1661  & 6:15:09.1,    71:08:12 & 60       & 0.75 &  20.8$\pm$0.2 & 0.042$\pm$0.005 & 19.7$\pm$0.2 & 0.066$\pm$0.003 \\
NGC 783          &  6$\times$600  & 1972  & 2:01:06.6,    31:52:57 & 43       & 0.75  &  19.9$\pm$0.3 & 0.062$\pm$0.010 & 18.4$\pm$0.3 & 0.110$\pm$0.005 \\
NGC 2500        &  7$\times$600  &  2974 & 8:01:52.7,    50:44:13 & 60       & 0.92  & 20.4$\pm$0.1 & 0.035$\pm$0.001 & 19.1$\pm$0.1 & 0.057$\pm$0.001  \\
\enddata
\tablenotetext{a}{Fits to $\mu_{V,0} = \mu_{\rm cen} + b R {\rm(arcs)}$ interior and exterior to \rbr.}
\end{deluxetable}

\begin{deluxetable}{lccc}
\tabletypesize{\scriptsize}
\tablecaption{FUV interior and exterior to \rbr\ \label{tab-ratsspirals}}
\tablewidth{0pt}
\tablecolumns{4}
\tablehead{
\colhead{}
& \colhead{}
& \multicolumn{2}{c}{log Interior/Exterior\tablenotemark{b}} \\
\colhead{Galaxy}
& \colhead{Break Type\tablenotemark{a}}
& \colhead{FUV/Area}
& \colhead{FUV/$V$}  \\
}
\startdata
KUG 0210-078 & II        &  $0.43\pm0.06$  &   $-0.09\pm0.01$ \\
NGC 3840        & II        &  $0.49\pm0.17$  &   $0.13\pm0.01$ \\
UGC 3422        & II        &  $0.18\pm0.38$  &   $-0.12\pm0.02$ \\
NGC 783          & II        &  $0.64\pm0.11$  &   $-0.06\pm0.01$ \\
NGC 2500        & II        &  $0.73\pm0.09$  &   $0.14\pm0.00$ \\
\enddata
\tablenotetext{a}{Type of surface brightness profile break in the $V$-band. 
''II'' refers to a downward break, and ''III'' to an upward bend.}
\tablenotetext{b}{The FUV flux is normalized by the area over which it is measured, ``FUV/Area,'' or by
the $V$-band flux measured over the same area, ``FUV/V.'' The ratio that is given is FUV/Area or FUV/$V$
measured interior to the surface brightness profile break to that measured exterior to the break.}
\end{deluxetable}

\end{document}